\documentclass[12pt]{JHEP3}
\usepackage{epsfig,latexsym,amssymb,cite}
\usepackage{amsmath}
\usepackage{graphicx}
\usepackage{bm}
\newcommand{\be}{\begin{equation}}
\newcommand{\ee}{\end{equation}}
\newcommand{\bea}{\begin{eqnarray}}
\newcommand{\eea}{\end{eqnarray}}
\newcommand{\bem}{\begin{multline}}
\newcommand{\eem}{\end{multline}}
\newcommand{\beg}{\begin{gather}}
\newcommand{\eeg}{\end{gather}}

\newcommand{\ben}{\begin{eqnarray*}}
\newcommand{\een}{\end{eqnarray*}}

\graphicspath{{Figs/}}
\title{ Heavy Ion Collisions with Transverse Dynamics from Evolving AdS Geometries}
\author{Anastasios Taliotis \\
\vspace{0.1in}
Department of Physics, The Ohio State University, Columbus,
OH 43210, USA \vspace{0.1in}
\\~~\\ 
E-mail addresses: \email{taliotis@mps.ohio-state.edu}
\vspace{0.1in}
}
\date{5$^{th}$ ofAugust, 2010}
\abstract{Currently there exists no known way to construct the Stress-Energy Tensor $(T_{\mu \nu})$ of the produced medium in heavy ion collisions at strong coupling from purely theoretical grounds. In this paper, some steps are taken in that direction. In particular, the evolution of $T_{\mu \nu}$ at strong coupling and at high energies is being studied for early proper times $(\tau)$. This is achieved in the context of the AdS/CFT duality by constructing the evolution of the dual geometry in an AdS$_5$ background. Improving the earlier works in the literature, the two incident nuclei  have an impact parameter $b$ and a non-trivial transverse profile. The nuclear matter is modeled by two shock waves corresponding to a non-zero five dimensional bulk Stress-Energy Tensor $J_{MN}$. An analytic formula for $T_{\mu \nu}$ at small $\tau$ is derived and is used in order to calculate the momentum anisotropy and spatial eccentricity of the medium produced in the collision as a function of the ratio $\frac{\tau}{b}$. The result for eccentricity at intermediate $\frac{\tau}{b}$ agrees qualitatively with the results obtained in the context of perturbation theory and by using hydrodynamic simulations. Finally, the problem of the negative energy density and its natural connection to the eikonal approximation is discussed.}

\preprint{}
\begin{document}

\section{Introduction}

There has been strong evidence from RHIC that the Quark-Gluon Plasma ({\bf QGP}) created in heavy ion collisions goes through a strongly coupled phase \cite{Kolb:2000sd} -\cite{Shuryak:2006se}. In particular, hydrodynamic simulations which describe the data successfully, require small shear viscosity compared to entropy density. This fact sends the message that very early after the collision the produced medium exhibits a strongly coupled behavior. Hydrodynamics simulations also use a short thermalization time of the order of 1 fm/c \cite{Kovchegov:2007pq}.

There have been theoretical efforts trying to obtain thermalization and more importantly isotropization at  early times within the context of perturbative QCD ( {\bf pQCD}) \cite{Baier:2000sb} - \cite{Kovchegov:2005ss}. While research in this direction is still under way, there has not been a satisfactory explanation of the (early) thermalization in the framework of perturbation theory. Nonetheless, perturbation theory is not only a powerful tool for describing collisions but also an essential one. The reason for this is due to the fact that the produced system starts out being weakly coupled and described in the Color Glass Condensate ({\bf CGC}) framework \cite{Blaizot:1987nc} -\cite{Krasnitz:2002mn} and subsequently becomes strongly coupled and thermalizes; the process takes place over times of the order of 1 fm/c. As a result, we should look for alternative methods to pQCD in order to describe the intermediate stages of heavy ion collisions.

Fortunately such a tool that may be applied to strongly coupled theories exists: The anti-de-Sitter /Conformal Field Theory correspondence ({\bf AdS/CFT}) was discovered in 1997 by Maldacena \cite{Maldacena:1997re} and was soon quantified by Witten \cite{Witten:1998qj}. Since then, its validity has passed many physics tests while it has been widely applied in problems related to QCD (and recently to condensed matter physics). Many observables or interesting
 quantities in QCD have been calculated and many processes have been studied within the framework of AdS/CFT. The list includes applications of AdS/CFT to the heavy quark potential \cite{Burnier:2009bk} -\cite{Brandhuber:1998bs}, to the evaluation of scattering amplitudes, in Deep Inelastic Scattering and to small-x physics  \cite{Marquet:2010sf} -\cite{Kovchegov:2009yj}, to parton motion in the plasma, energy loss and radiation \cite{Horowitz:2009pw} -\cite{Beuf:2008ep},
 heavy ion collisions, QGP and the hydrodynamic behavior of the produced medium \cite{Policastro:2001yc} - \cite{Kovchegov:2007pq} and \cite{Kajantie:2006hv} - \cite{Nastase:2005rp} and many others. Regarding heavy ion collisions, the question of whether a thermal medium is produced may be attacked by looking at trapped surfaces in the gravity side. This has been done in \cite{Gubser:2008pc} - \cite{Giddings:2004xy} and references therein, and the method has been adapted from earlier work on general relativity. 
We stress that AdS/CFT does not deal with QCD but with  ${\cal N}=4$ SYM theory with an $SU(N)$ gauge group snd with $N$ being large. In  \cite{Panero:2009tv} one may find the values of several thermodynamic quantities computed by applying lattice methods with finite $N$ at large coupling and comparisons with results obtained from AdS/CFT \cite{Gursoy:2007cb,Gursoy:2007er,Gursoy:2008bu,Kiritsis:2009hu}. There are more differences between the two theories and certainly many similarities, especially at the deconfining limit. Reference \cite{Edelstein:2009iv} exposes these matters and reviews the progress in the field, starting from 't Hooft's large $N$ field theories \cite{'tHooft:1973jz} and building up the discovery of the gauge/gravity correspondence and its applications to problems related to gauge theories and QCD.

In this paper, we are interested in studying the problem of heavy ion collisions in the context of the gauge-gravity correspondence.  In particular, we are interested in calculating the Stress-Energy {\bf(SE)} tensor of the produced medium and related quantities along with the way they evolve with time. This is achieved by finding the time evolution of the dual geometry in an AdS background since according to the duality, the metric tensor on the gravity side is dual to the (expectation value of the) {\bf SE} tensor in the gauge theory side. We assume that the nuclei that collide have non-trivial transverse profiles and that the collision is not central but there is a nonzero impact parameter (b) involved \footnote{In fact, the impact parameter is required, for otherwise one has to face violation of conservation of the bulk stress-energy tensor (see (\ref{Jcon})) and ultraviolet ({\bf UV}) infinities (see subsection \ref{con}).}. To the best of our knowledge, this is the first time in the literature where in the framework of the AdS/CFT correspondence modeling of heavy ion collisions with transverse dynamics is taken into account in order to calculate the evolution of the {\bf SE} tensor of the produced medium. The collision is assumed to take place at large 't Hooft coupling \footnote{Although this is not the case in real QCD as at the very early times, as has been already mentioned, pQCD seems to describe collisions satisfactorily.} and our analysis applies at early proper times in a way that is fully explained and quantified.

 The problem from the gravitational perspective is formulated as an initial value problem where the dual geometry is known in some time interval (in contrast to \cite{Kovchegov:2007pq}, \cite{Janik:2005zt,Janik:2010we}). This initial geometry is chosen as two shock waves moving towards each other and represents the nuclear matter to be collided. The evolution of the geometry, as determined by Einstein's equations, maps onto the evolution of the produced matter of the gauge theory in a way which is made precise by the AdS/CFT correspondence. The paper \cite{deHaro:2000xn} quantifies the relation between the {\bf SE} tensor of the gauge theory and the metric tensor on the gravity side. A discussion which elucidates this issue through particular examples may be found in \cite{Kovchegov:2007pq} and \cite{Janik:2005zt}. Our work has partial overlap with \cite{Albacete:2009ji}-\cite{Janik:2006ft} and may be regarded as an improvement of \cite{Albacete:2008vs} which in turn was motivated by \cite{Kovchegov:2007pq}. All these papers employ the results of \cite{deHaro:2000xn}. We also note that the literature about shock waves in gravity is very rich; references \cite{DEath:1992hb} - \cite{Sfetsos:1994xa} include only a subset of it.


\vspace{0.3in}

We organize our paper as follows.

\vspace{0.3in
}
 In section \ref{setup} we set up the problem, explain the precise way that the problem of heavy ion collisions is mapped onto a gravity problem in the context of AdS/CFT and state how the {\bf SE} tensor of the gauge theory can be computed from the metric of the dual theory.

 In section \ref{trp} we talk about the transverse part of the nuclear profiles and how these may be modeled in terms of gravitational shock waves which correspond to a non-zero bulk (gravity) {\bf SE} tensor. We argue that in order to have a well defined and non-trivial  metric, meaning that it is finite both in the UV and the infrared ({\bf IR}) in AdS bulk, a bulk source is necessary (compare with \cite{Beuf:2009mk}). This is analogous to classical electrodynamics where the Laplace equation has nontrivial solutions which are finite at both the origin and at infinity only in the presence of a charge density. In particular we derive the Green's function (\ref{g8}) of the scalar operator for AdS$_5$ which has the desired boundary conditions.

In section \ref{Jcor}, we take into account back-reactions in the gravity description in order to have a conserved and therefore a well defined bulk {\bf SE} tensor (see (\ref{Jmn2}) and (\ref{Jmnc})). Here, we take into account that the first bulk source moves in the gravitational field of the second and vice versa. This alters the trajectories of the bulk sources and induces a time evolution in the matter density in the bulk as figure \ref{BTen} suggests. In a diagrammatic approach, these corrections to the initial bulk {\bf SE} tensor would correspond to the diagrams of figure \ref{Self_Int}.

 Section \ref{FEq} includes Einstein's equations and the way these equations may be decoupled in order to be solved perturbatively. The program is to expand the field equations in the AdS$_5$ background in the presence of the two shock waves and solve them order by order in the strength of the shock waves. At each step (order of the expansion) we make sure that the bulk {\bf SE} tensor is conserved (see section \ref{Jcor}). The first correction to the metric corresponds to the diagrams of figures \ref{interaction} and \ref{Self_Int}. Obviously, according to the dictionary of the AdS/CFT, this correction to the metric maps onto a correction of the initial {\bf SE} tensor of the colliding nuclear matter. Finding this correction is the ultimate goal of our work. 

 In section \ref{Cal} we exhibit the main details of the calculation and state the main result of this paper, that is the formula giving the gauge theory {\bf SE} tensor of the matter produced in a collision. This is given by (\ref{SE}) while the details of the calculation are left for the two Appendices at the end of the paper.

 Finally, in section \ref{VIC} we discuss the area of validity of our approximations and quantify the meaning of the fact that our approximation applies for early times. This is done in subsection \ref{va}.

In subsection \ref{IS} we estimate the energy density and the transverse pressures at certain kinematical regions (see (\ref{eted}) and (\ref{lted})) where our analytic formula (see (\ref{SE}) and (\ref{TMN})) simplifies to give compact expressions for the {\bf SE} tensor \footnote{Certainly our analytic formula is (more) accurate but unfortunately rather complicated: we find it useful to investigate particular limits of it.}. In particular, for the regions $I$, $I$$^{'}$ and $III$, $III$$^{'}$ of figure \ref{re} we find that the energy density  behaves as in (\ref{eted}) and (\ref{lted}) respectively. The result for  regions $III$, $III$$^{'}$  exhibits a similar logarithmic dependence on proper time as the results of \cite{Lappi:2003bi} - \cite{Fries:2006pv} although our result (at central rapidities) has an additional overall factor of proper time squared. In \cite{Lappi:2003bi} - \cite{Fries:2006pv} the results are obtained in the context of pQCD. 

In subsection \ref{Eccsec} we use the approximate results of previous subsection in order to derive approximate formulas for the momentum anisotropy and the spatial eccentricity of the produced medium. In the two particular limiting cases that we have investigated in subsections \ref{IS}, the transverse pressure components of the {\bf SE} tensor turn out to be equal. Therefore we conclude that the momentum anisotropy (see (\ref{man})) should be close to zero. In addition, for one of the two cases we estimate the spatial eccentricity as a function of the ratio $\frac{\tau}{b}$ for intermediate  proper times $\tau$ (see (\ref{EcIII}) and figure \ref{ecfig}). Our result is qualitatively consistent with the results of \cite{Lappi:2006xc,Jas:2007rw} obtained from pQCD and the results of \cite{Kolb:2000sd} and \cite{Kolb:2003dz,Kolb:2002cq} obtained by using hydrodynamic simulations. Our analytic formula for the {\bf SE} tensor, equation (\ref{SE}), allows us to extend these results to a broader kinematical region than the two aforementioned limiting cases (of subsection \ref{IS}); this is left for a future project.

In subsection \ref{con} we summarize our conclusions: The evolution of the gauge theory {\bf SE} tensor is constrained by causality \footnote{Certainly causality is already built to the dynamics of the problem.} (see figure \ref{re} and equation (\ref{TMN})) in an intuitive way. Causality basically separates the evolution into three \footnote{Totally there are six different areas but only three of them are not trivial. The other three may be obtained from the non-trivial regions by using mirror symmetry.} areas which are shown in figure \ref{re} (see also equation (\ref{TMN})). Mathematically this is due to the presence of retarded propagators which induce $\theta$-functions, and which in turn separate space-time according to figure \ref{re}: the {\bf SE} tensor at any given point of space evolves according to whether the propagation of the signal from the center of each nucleus (or both nuclei) has enough proper time to reach the point under consideration. Since our approach is generally applicable for early (proper) times we can not unfortunately test for thermalization. In the best case one may follow the steps outlined in subsection \ref{con}, conclusion 2., which provide a necessary but not sufficient condition whether a thermal medium is produced. The method, may also give an estimation of the times where thermalization occurs (if it does). We also illustrate how the momentum anisotropy and the spatial eccentricity may be obtained from the expression of $T_{\mu \nu}$ (equation (\ref{SE})) we have derived applying AdS/CFT and we compare our result with the results of the literature obtained by using pQCD or  by hydrodynamical methods.  In addition we argue about the necessity of an impact parameter and also about the fact that the bulk sources do contribute to the metric even at space-time points away from them. Furthermore, motivated by our result for the {\bf SE} tensor, we propose how the energy density should evolve in more realistic cases. We construct a simple phenomenological model taking into account the Woods-Saxon profile of a nucleus at rest. We conjecture that the energy density at early times and close to the center of the collision should behave as in (\ref{WS}). Finally, we discuss the problem of negative energy density and how it is associated with the eikonal approximation in collisions.

\section{Setting up the problem}\label{setup}

We suppose that we have a single nucleus moving with the speed of light and use the AdS/CFT correspondence in order to model this in terms of a dual metric. We assume that the nucleus moves along decreasing $x^3$ and has a transverse ({\bf SE} tensor) profile possessing azimuthal symmetry. The metric that realizes this situation corresponds to a  metric \footnote{As we see the nucleus is dual to a  in gravity; hence we will use both notions interchangeably.} and a simple choice \footnote{In section \ref{trp} we will show how to construct more complicated  metrics. These correspond to different nuclei profiles.} may be taken to be \footnote{We choose the signature of the metric to be $(-,+,+,+,+)$.}

\begin{align}\label{s1}
ds^2 \, = g_{MN}dx^M dx^N=\, \frac{L^2}{z^2} \, \left\{ -2 \, dx^+ \, dx^- + t_1
(x^+,x^1,x^2) \, z^4 \, d x^{+ \, 2} + d x_\perp^2 + d z^2 \right\}, \hspace{0.01in}x^M= x^\mu,z.
\end{align}
Here $d x_\perp^2 = (d x^1 )^2 + (d x^2)^2$ is the transverse metric and $x^\pm = (t \pm x^3) / \sqrt{2}$ where $x^3$
is the collision axis and $t$ the time axis. With $x^M$ we include the four Minkowski components $x^{\mu}=x^1,x^2,x^3$ and $x^0=t$ and the component $z$ which is the fifth dimension of AdS$_5$ with $L$ being its radius. Now, $t_1$ of (\ref{s1}) cannot be an arbitrary function but should be constrained by Einstein's equations in AdS space. In the presence of matter these equations may be cast in the convenient form
\begin{align}\label{Ein}
R_{MN} + \frac{4}{L^2} \, g_{MN} \, = \kappa_5^2
\left( J_{MN} - \frac{1}{3} \, g_{MN} \, J \right)
\end{align}
with 
\begin{align}\label{BT}
J = J_M^{\ M} = \, J_{MN} \, g^{MN}\hspace{0.35in}\kappa_5^2=8\pi G_5.
\end{align}

In equation (\ref{Ein}), $R_{MN}$ is the Ricci tensor, $g_{MN}$ is the metric tensor in five dimensions and $J_{MN}$ the bulk {\bf SE} tensor \footnote{The bulk {\bf SE} tensor should not to be confused with the $T_{\mu \nu}$ tensor of the produced matter in the four dimensional  gauge theory.}. $G_5$ is the Newton's constant in five dimensions with $[G_5]$=-3. One can check that (\ref{s1}) solves (\ref{Ein}) exactly with a vanishing bulk tensor provided that 

\begin{align}\label{do}
\frac{1}{2} \left(\frac{3}{z} \partial_z-\partial_z^2-\nabla_{\perp}^2 \right)z^4 t_1(x^+,x^1,x^2)=0,
\end{align}
where $\nabla^2_{\perp}$ is the 2 dimensional Laplacian in the transverse plane. This means that the following condition should be satisfied

\begin{align}\label{Lap}
-\frac{1}{2}z^4\nabla^2_{\perp}t_1(x^+,x^1,x^2)=0.
\end{align}

We are interested in rotationally invariant eigenfunctions for this differential operator. However, the only choice \footnote{Apart from a trivial constant (see \cite{Albacete:2009ji,Albacete:2008vs,Grumiller:2008va}).} is the $\log(x_1^2+x_2^2)$ function which induces a delta function on the right-hand side. We interpret the appearance of such a delta function as a (nonzero transversely) localized $J_{MN}$. In particular, we choose the  profile to be

\begin{align}\label{t1}
t_1(x_+,x_1,x_2)=-\mu \hspace{0.01in} \log(k r)\delta(x_-) ,\hspace{0.3in} r=\sqrt{(x^1)^2+(x^2)^2}.
\end{align}
In this equation $k$ serves as a cutoff and its precise meaning will become clear in the next section while $\mu$ has mass dimensions $[\mu]$=3. As we will see shortly, $\mu$ is associated with the (expectation value of the) {\bf SE} tensor of the corresponding nucleus in the gauge theory and will serve as an expansion parameter of the metric. Taking into account that

\begin{align}\label{nablat}
\nabla^2_{\perp}\log(k r) = 2\pi \delta (\vec{r}) \hspace{0.3in} \vec{r}=(x^1,x^2),
\end{align}
we realize that there is a nonzero right-hand side for equations (\ref{Ein}) associated with a five dimensional bulk {\bf SE} tensor. It turns out that this bulk tensor is given by \footnote{In equation (\ref{J1}) we assume that the center of the  is shifted away from the origin by distance b along the positive $x^1$ axis.}

\begin{align}\label{J1}
J^{(1)}_{++} \, = \frac{\pi \mu}{\kappa_5^2}z^4 \, \delta (x^1-b)\delta (x^2)\delta (x^+).
\end{align}
All other components are zero. The presence of the superscript $(^{(1)})$ on $J_{++}$ is to highlight that it is of first order in the parameter
 $\mu$ \footnote{The superscripts emphasize the number of times the source $t_i$ (i=1,2 assuming we have two sources) appears. Generally in the object $A^{(n)}$ there exists the product  $t_1^k t_2^{n-k}$ or any linear combination of differentiation/integration of $t_1$ and $t_2$ with respect to their arguments $x^{\mu}$.}.
 In this special case of a single , the first order solution happens to be the exact solution to all orders. In a diagrammatical approach, this solution would correspond to the diagram of figure \ref{vertex}. It represents the measurement of the gravitational field at point $x^M$, which, loosely speaking, is created by a single graviton emission from the bulk source of equation (\ref{J1}) with effective coupling proportional to $\mu \log(k r)$.
 
\FIGURE{\includegraphics [scale=0.45]{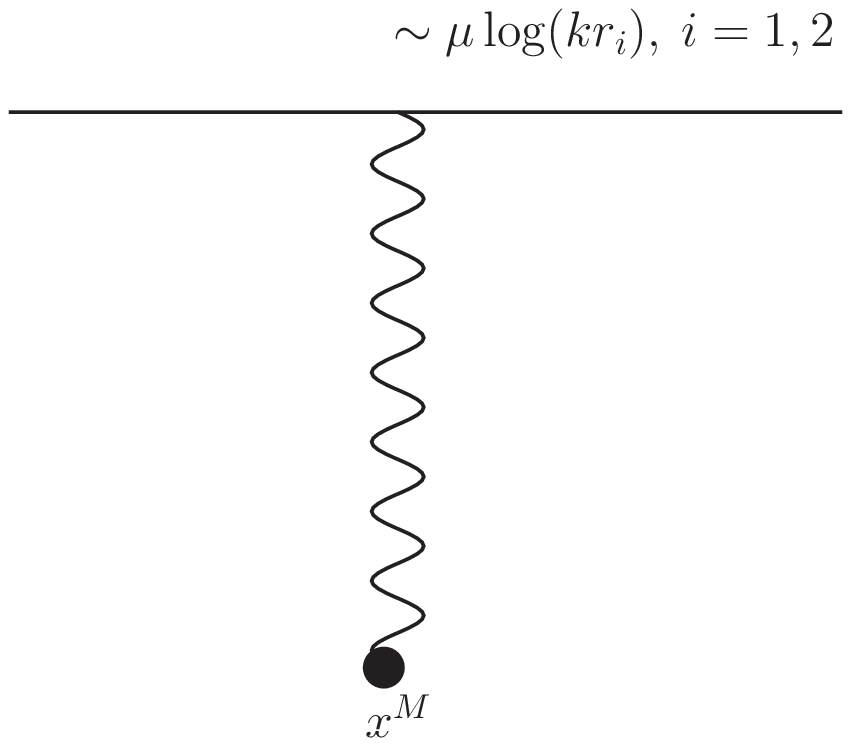}
  \caption{The  solution: a graviton is emitted from the bulk source with coupling $\mu \log(kr)$ which is measured at the point $x^{M}$. This is a very special case where a single graviton exchange between the source and the bulk happens to be an exact solution to the nonlinear Einstein's equations.}
  \label{vertex}
  }
This bulk source is point-like along $x^1$, $x^2$ and $x^3$, has infinite extent along the $z$ direction and moves towards decreasing $x^3$. It represents a one-dimensional ``string'' of matter moving along the $x^-$-axis. It is important to note that this bulk tensor is (covariantly) conserved as it should be. On the other hand, the boundary physics is obtained using holographic renormalization \cite{deHaro:2000xn}. We find that the {\bf SE} tensor corresponding to the metric (\ref{s1}) has only one non-zero
component. To see this, we use 


\begin{align}\label{hr}
\langle T_{\mu\nu}(x^{\mu}) \rangle \, = \, \frac{L^3}{4 \, \pi \, G_5} \, \lim_{z \rightarrow 0} \frac{g_{\mu\nu}(x^{\mu},z)-\eta_{\mu \nu}}{z^4} , \hspace{0.3in}G_5=\frac{\pi L^3}{2 N_c^2}.
\end{align}
where we assume that the metric is written in the Fefferman-Graham coordinate system

\begin{align}\label{FG}
ds^2 \,& = \frac{L^2}{z^2} \bigg\{ 
g _{\mu \nu}(x^{\kappa},z)dx^{\mu}dx^{\nu} + d z^2 \bigg\}.
\end{align}
Combining previous two equations we deduce that

\begin{align}\label{GT1}
\langle T_{++}(x^{\mu}) \rangle \, = \, \frac{L^3}{4 \, \pi \, G_5} \, \lim_{z \rightarrow 0} \frac{g_{++}(x^{\mu},z)}{z^4} \, =\frac{N_c^2}{2\pi^2} t_1= -\frac{N_c^2}{2\pi^2} \mu \log(k r) \delta (x^+).
\end{align}
In both equations (\ref{hr}) and (\ref{GT1}), $N_c$ is the number of colors. We note that $T_{++}(x^\mu)$ is positive for $k r \leq 1$ and therefore at first sight $1/k$ gives the maximum distance where the {\bf SE} tensor of the nuclear matter is physical. In fact, this logarithmic (transverse) profile of the nucleus is a part of a more complicated profile which is both positive definite for all $r$ and reduces to the one of (\ref{GT1}) for small distances. Moreover, the corresponding metric is minimal in the sense of being the simplest one that can simultaneously capture some of the transverse dynamics of heavy ion collisions and allow for an analytical approach to the problem. We will give a detailed justification of these claims in the next section.

\begin{figure}
\centering
\includegraphics[scale=0.5]{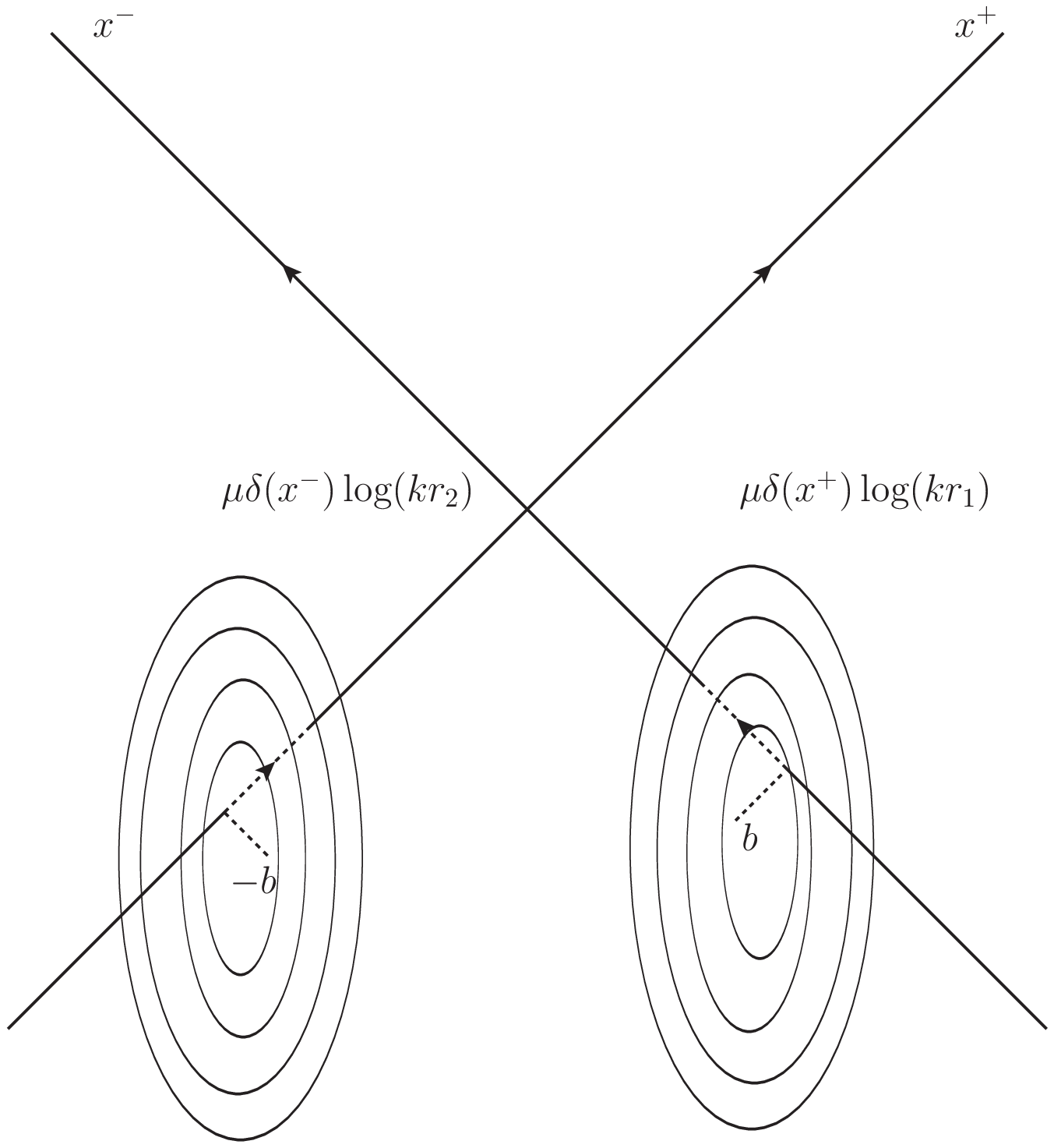}
 \caption{The two nuclei moving along $x^{\pm}$ axis and colliding at the origin. Along with them, they drag a perpendicular energy-momentum density which is constant along the circular lines.}
  \label{offcenter}
 \end{figure}

Having defined all the necessary ingredients we now proceed to the main part of the setup. We want to superimpose two such shock waves whose sources are two one dimensional distributions of matter in the bulk. We want to collide them at a non-zero impact parameter and hence study the problem within the classical theory of gravity. Therefore, $J_{MN}$ has in addition to (\ref{J1}) the symmetric part 

\begin{align}\label{J2}
J^{(1)}_{--} \, = \frac{\pi \mu}{\kappa_5^2} z^4\delta (x^-) \, \delta (x^1+b)\delta (x^2)
\end{align}
which creates a second . In terms of the gauge theory, this would correspond to colliding two nuclei in an off center process. Figure \ref{offcenter} represents the four dimensional picture, right before the collision of the two nuclei. Following \cite{Albacete:2009ji},\cite{Albacete:2008vs} and working with the  Fefferman-Graham coordinate system, the metric that describes the process should look like

   \begin{figure}
\centering
\includegraphics[scale=0.55]{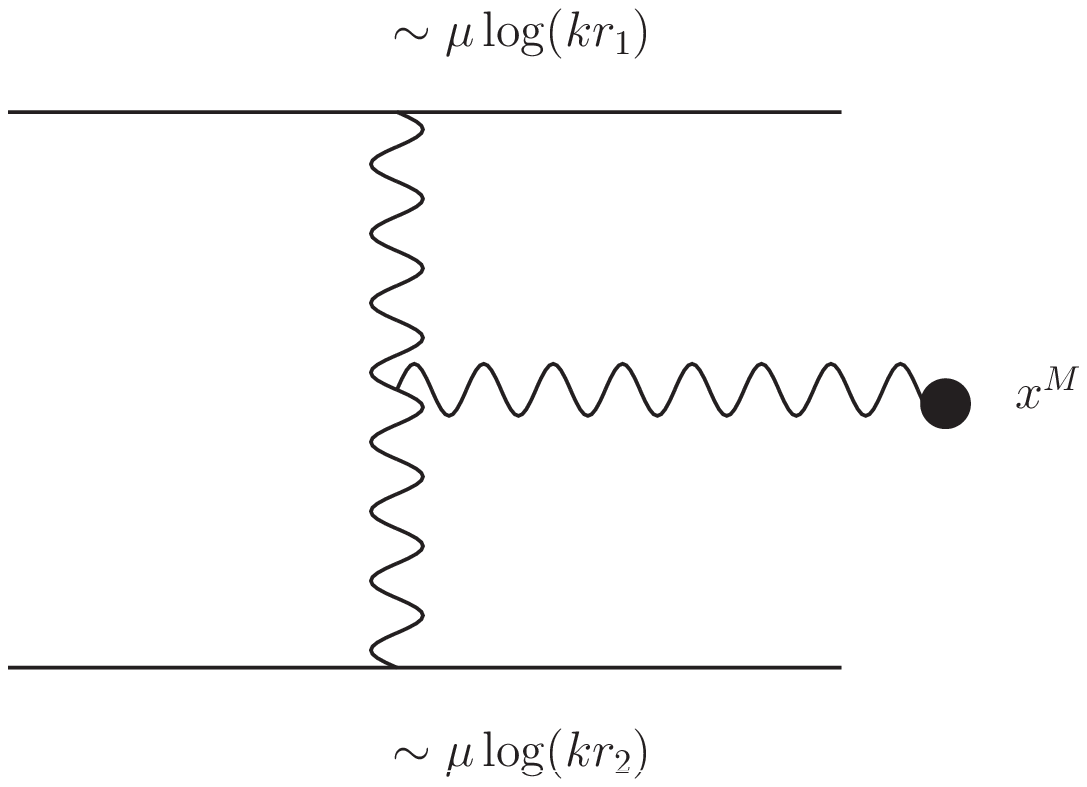}
 \caption{The Feynman diagram that represents the $\mu^2$ correction of the metric:  It represents the first nontrivial correction to (2.13). It shows  how the two metrics each of which like (2.1) merge. The gravitational field is measured at the point $x^{M}$.}
  \label{interaction}
 \end{figure}

\vspace{0.3in}

\begin{align}\label{s12}
ds^2 \, &= \frac{L^2}{z^2} \bigg\{ -2 \, dx^+ \, dx^- + d x_\perp^2 + d z^2+ t^{(1)}_1(x^+,x^1-b,x^2) \, z^4 \, d x^{+ \, 2} \notag\\&
 + t^{(1)}_2(x^-,x^1+b,x^2) \, z^4 \, d x^{- \, 2}
+ \theta(x^+)\theta(x^-)g^{(2)} _{\mu \nu}(x^{\kappa},z)dx^{\mu}dx^{\nu}   + \ldots       \bigg\}, \notag\\&
 t^{(1)}_{1,2}(x^1 \mp b,x^2)= -\mu \log\left (k \sqrt{(x^1 \mp b)^2+(x^2)^2}\right) \delta(x^{\pm}).
\end{align}

\vspace{0.5in}
\hspace{-0.3in}The first three terms correspond to the empty AdS$_5$. The next two are of first order in $\mu$ and are created by the two lines of distribution of bulk matter that stretch along the $z$ direction. These objects move (initially) towards each other along $x^3$ and they have an impact parameter $2b$ along the $x^1$ axis as figure \ref{BTen} depicts.  As they are the sources of the two shock waves, they correspond to two vertex diagrams that look like the one in figure \ref{vertex}. This is a superposition of two metrics with each one looking like (\ref{s1}) while the gauge theory picture is shown in figure \ref{offcenter}. However, the non-linearities of the gravitational field require higher order terms. The second order corrections are explicitly displayed in (\ref{s12}) and they appear once the two shock waves cross each other in the forward light cone. This is precisely the meaning of the theta functions; they emphasize that the metric (\ref{s12}) solves Einstein's equations  exactly in the presence of both shock waves only for negative $x^{\pm}$. According to holographic renormalization \cite{deHaro:2000xn} and equation (\ref{GT1}), the additional terms of the metric that appear in the forward light cone correspond to the matter of the produced medium in gauge theory after the nuclear collision. The main work of this paper is to calculate them to order $\mu^2$, that is find $g_{\mu \nu}^{(2)}$ and determine $T_{\mu \nu}$ of the gauge theory. The second order correction in $\mu$ of $g_{\mu \nu}$ corresponds to the diagram of figure \ref{interaction}.


\section{Choosing the transverse profiles of the initial nuclear matter }\label{trp}

In this section we justify the choice of the profiles of the shock waves which look like $t_{1}$  of equation (\ref{t1}). According to equation (\ref{t1}) and (\ref{nablat}) we have that 

\begin{align}\label{g1}
\left(\frac{3}{z} \partial_z-\partial_z^2-\nabla_{\perp}^2 \right)(-z^4 \log(k r)) = 2\pi z^4\delta(\vec{r})
\end{align}

We will see this differential equation from a different perspective. We will search for the Green's function of the operator (\ref{do}) and then use it in order to determine $t_1$ by assuming that the source of this operator behaves as $\sim z^4 \delta(\vec{r})$ . We need the function $G$ which satisfies

\begin{align}\label{g2}
\left(\frac{3}{z} \partial_z-\partial_z^2-\nabla_{\perp}^2 \right) G = \delta (z-z')\delta(\vec{r}-\vec{r}_0)
\end{align}
and which is finite as $z\rightarrow 0,\infty$. We will see that the solution $t_1 \sim z^4 \log(k r) $ is only a part of a more complicated  metric and we will also associate $k$ with an IR cutoff of the fifth AdS coordinate z. 
Searching for solutions of the form

\begin{align}\label{g3}
G=\int \frac{d^2q}{(2 \pi)^2} e^{i \vec{q}(\vec{r}-\vec{r'})}A(\vec{q},z,z')
\end{align}
we find that A should satisfy

\begin{align}\label{g4}
\left(\frac{3}{z} \partial_z-\partial_z^2+q^2 \right) A = \delta (z-z').
\end{align}
It turns out that the two independent solutions for A are $z^2 K_2(qz)$ and $z^2 I_2(qz)$ where $K_2$ and $I_2$ are Bessel functions.
Now, the boundary conditions together with continuity of G require that A should behave like

\begin{align}\label{g5}
A \sim \theta(z'-z)z^2 K_2(qz') I_2(qz)+\theta(z-z') z^2 K_2(qz) I_2(qz').
\end{align}

In order to determine the overall coefficient, we substitute (\ref{g5}) in (\ref{g4}). The result is

\begin{align}\label{g6}
A = \frac{z^2}{z'} \left[ \theta(z'-z) K_2(qz') I_2(qz)+\theta(z-z') K_2(qz) I_2(qz') \right].
\end{align}

The next step in to plug (\ref{g6}) in (\ref{g3}) and perform the angular integration. This yields to

\begin{align}\label{g7}
G= \frac{z^2}{2 \pi z'} \int_{0}^{\infty} dq \hspace{0.01in}q J_0(q|\vec{r}-\vec{r}'|) \left( \theta(z'-z) K_2(qz') I_2(qz)+\theta(z-z') K_2(qz) I_2(qz') \right).
\end{align}
These integrals \footnote{Both integrals converge as $q\rightarrow \infty$ because of the presence of the $\theta$ functions. The values of the integrals turn out to be the same under $z\leftrightarrow z'$ and hence independent of the ordering of $z$ and $z'$.} are tabulated in \cite{GR} and the answer involves the associated Legendre functions of the second kind $(Q_{\nu}^{\mu})$. We find

\begin{align}\label{g7b}
G= \frac{z}{2 \pi z'^2}\frac{1}{2^{\frac{5}{2}} \sqrt{\pi}} e^{i \frac{\pi}{2}} (u^2-1)^{-\frac{1}{4}} Q_{1/2}^{3/2}(u), \hspace{0.1in}u=\frac{|\vec{r}-\vec{r'}|^2+(z-z')^2}{2zz'}+1.
\end{align}
The function $Q_{1/2}^{3/2}$ may be simplified even further \cite{GR} in terms of a hypergeometric function. The final result is \footnote{The fact that G of (\ref{g7}) is not symmetric under $z \leftrightarrow z'$ is because the differential operator (\ref{g2}) is not Hermitian.}

\begin{align}\label{g8}
G(\vec{r},z;\vec{r'},z') =\frac{z}{16 \pi z'^2} \frac{1}{u^3} F(2,3/2,3,1/u^2) .
\end{align}

We may also cast last equation in the rather simple form

\begin{align}\label{g8b}
G(\vec{r},z;\vec{r'},z')&=\frac{1}{16 \pi z'^3} \Bigg \{-4|\vec{r}-\vec{r'}|^2+4z^2\left(\frac{2 z'^2}{\sqrt{|\vec{r}-\vec{r'}|^4+(z^2-z'^2)^2+2 r^2(z^2+z'^2)}}-1\right)\notag\\&
+4 \left(-z'^2+\sqrt{|\vec{r}-\vec{r'}|^4+(z^2-z'^2)^2+2 r^2(z^2+z'^2)}\right) \Bigg \}.
\end{align}
This result agrees with \cite{Gubser:2008pc}, \cite{Gubser:2009sx} and \cite{Hotanaka:1993} and it was obtained more elegantly by using the symmetries of the AdS$_d$ space and reducing the problem to solving a differential equation of a single variable $u$. 

Now we will make use of the Green's function we derived in order to answer the question we posed at the beginning of this section. We wanted to find the solution of the following differential equation

\begin{align}\label{g9}
\left(\frac{3}{z} \partial_z-\partial_z^2-\nabla_{\perp}^2 \right) h =2 \pi z^4 \delta(\vec{r}-\vec{r'})\delta(x^+).
\end{align}We have to take the product of the right-hand side of (\ref{g9}) with (\ref{g8b}) and integrate. The transverse integration is trivial. But then, one may observe that the integrand of the product of (\ref{g9}) with (\ref{g8b}) at large $z'$ behaves as $ \frac{z^4}{z'}$. The only available scale that fixes the dimensions is $r$ and hence the result should receive a $z^4\log(\frac{z'}{r})$ contribution from large $z'$. This suggests we should place an infrared cutoff $z_c$ on $z'$.

Equivalently, we may argue in a more physical way. We may alternatively ask for the solution of

\begin{align}\label{g9b}
\left(\frac{3}{z} \partial_z-\partial_z^2-\nabla_{\perp}^2 \right) h =2 \pi z^4 \theta(z_c-z) \delta(\vec{r}-\vec{r'})\delta(x^+)
\end{align}assuming that the bulk {\bf SE} tensor does not extend to infinity in the $z$ direction but the source in the bulk looks like a rod of finite length $z_c$. The exact result for $h$ for this bulk tensor is then 

\begin{align}\label{g9c}
\hspace{1in}h&= G \otimes 2\pi  z'^4 \theta(z_c-z') \delta(\vec{r}-\vec{r}')\delta({x^+})\notag\\&
= \frac{1}{8}\delta({x^+})  \Bigg\{-2(r^2+z^2)z'^2-z'^4+(r^2+3z^2+z'^2)\sqrt{r^4+(z^2-z'^2)^2+2 r^2(z^2+z'^2)}\notag\\&
+4z^4 \log \left(2\left( r^2-z^2+z'^2+\sqrt{r^4+(z^2-z'^2)^2+2 r^2(z^2+z'^2)}\right) \right) \Bigg \}\Bigg|^{z_c}_{0}
\notag\\&
 =-z^4\delta({x^+})\left( \log(\frac{r}{z_c})+\frac{z^2-3 r^2}{2z_c^2} \right)+ O(\frac{1}{z_c^4}),
\end{align}
up to the term $-\frac{3}{4}z^4\delta(x^+)$ which we have absorbed in the cutoff $z_c$. Our approach to the problem is to work only with the logarithmic piece of (\ref{g9c}) assuming that $z_c$ is large enough. Thus, whatever results we will obtain using the shock waves of (\ref{s12}), should be thought of as arising from the (exact expression of the) shock waves of (\ref{g9c}), up to corrections of $O(\frac{1}{z^2_c})$. We stress that the logarithmic piece is also an exact solution of Einstein's equations with a bulk stress energy tensor that has the form of (\ref{J1}). This bulk tensor is different than the right-hand side of equation (\ref{g9b}) but the two agree in the limit where $z_c \rightarrow \infty$.

We thus observe in a direct way how the $z$ direction of AdS$_5$ plays a crucial role in the transverse behavior of the gauge theory {\bf SE} tensor \cite{Erlich:2005qh} - \cite{Grigoryan:2007wn}. In addition, as $r\rightarrow z_c^{+}$ the gauge theory {\bf SE} tensor of the  that corresponds to the exact expression for $h$, equation (\ref{g9c}), remains positive. In order to see this, we expand the exact solution of (\ref{g9c}) to order $z^4$ obtaining

\begin{align}\label{g9d}
h\Big | _{z^4}=\frac{1}{8}\delta(x^+) \left( 4\log(1+\frac{1}{x^2})-2\frac{2 x^2+3}{(1+x^2)^2} \right)z^4, \hspace{0.3in}x\equiv \frac{r}{z_c}\hspace{0.02in}.
\end{align}
Equation (\ref{g9d}) is (proportional to) the exact $T_{\mu \nu}$ of the gauge theory which corresponds to the exact  produced by a rod of length $z_c$ according to the right-hand side of (\ref{g9b}). It is monotonically decreasing to zero and is strictly positive for all $r$. It reduces to the approximate expression of (\ref{g9c}) when $z_c \rightarrow \infty$. It is now evident that we may identify $1/z_c$ with $\sim k$ of equations (\ref{t1}) and (\ref{s12}).

In addition to the above analysis, we may derive a similar conclusion by working from a different perspective. We may associate the (IR) cutoff of the parameter $q$ (see below) of (\ref{g7}) with some infrared scale, probably comparable with $\Lambda_{QCD}$. We therefore convolute $G$ of (\ref{g7}) with the right-hand side of (\ref{g9}) and perform the trivial $d^2 r'$ integration and also the $z'$ integration. We obtain

\begin{align}\label{10}
h= G \otimes 2\pi z'^4\delta(\vec{r}-\vec{r}')\delta({x^+})=z^4 \int_{0}^{\infty} dq \frac{J_0(q|\vec{r}-\vec{r}'|)} {q}\delta({x^+}).
\end{align}
We immediately encounter a singularity at $q=0$. In order to regulate the singularity, we place an IR cutoff on $q$. Now, by performing the $q$ integration we obtain the final result

\begin{align}\label{g11}
t_1=\frac{h}{z^4}&= \int_{k}^{\infty} dq \frac{J_0(q|\vec{r}-\vec{r}'|)} {q}\delta({x^+})\notag\\&
= \left \{-\log(k r) + \frac{ k^2 r^2}{8} F_{PFQ}\left(1,1;2,2,2;-\frac{k^2 r^2}{4}\right) + const. \right \}\delta({x^+})
\end{align}
where $ F_{PFQ}$ is a generalized hypergeometric function. Therefore, as long as we work in the kinematical region r, $ \tau \ll1/k$ we may ignore the hypergeometric piece of equation (\ref{g11}) and safely work with the logarithmic piece \footnote{ See section \ref{VIC} for details.}.

Working with the logarithmic piece of the transverse profile means that the energy density of the single beam moving along the light cone is infinite but integrable at the center of the beam and tends to zero logarithmically and radially outwards; it vanishes at $r=1/k$. This gives an additional indication that $1/k$ should be of the order of the nucleus size \cite{Erlich:2005qh} - \cite{Grigoryan:2007wn}, that is 

\begin{align}\label{kL}
k \sim \frac{\Lambda_{QCD}}{A^{\frac{1}{3}}}
\end{align}
where $A$ is the atomic number. Under this assumption, the total energy of the beam can be estimated by

\begin{align}\label{Ene}
E \sim -  \mu \int_{-\infty}^{\infty} dx^+ \int_{0}^{1/k} dr \delta (x^+) r \log(k r) \sim  \mu \frac{A^{\frac{2}{3}}}{\Lambda^2_{QCD}}.
\end{align}
Taking into account that the energy of the matter moving along $x^-$ should be proportional to the $p^-$ momentum of each nucleon of the nucleus and hence $E \sim p^- A$, we deduce that

\begin{align}\label{mu}
\mu\sim p^- \Lambda^2_{QCD} A^{\frac{1}{3}}
\end{align}
which is in agreement with \cite{Albacete:2008vs,Albacete:2009ji}.
What is more is that this choice of a nuclear profile is very simple in the sense that the $z$ and the transverse coordinate dependence $r$ of the  factor. This will simplify the calculations significantly although they are still not trivial. We note that we were able to construct nuclear profiles that behave like $\sim \frac{1}{r^{n}}$ with $n$ being an integer. These correspond to dual metrics that are much more involved and the calculations would have been harder. Finally we find it attractive to work with a metric that replaces the $\sim -z^4 \log (k r)$ profile with (\ref{g8}). This metric gives a nuclear profile $\sim \frac{1}{(r^2+z'^2)^3}$ as in \cite{Gubser:2008pc} with $z'$ being an arbitrary constant. The advantage of such a  metric is that it gives a point-like bulk (five dimensional) {\bf SE} tensor and the nuclear profile is everywhere positive. Unfortunately this metric is relatively involved and makes calculations hard. We leave the investigation of such a case for a future project.


\section{Back-Reactions}\label{Jcor}


\subsection{Corrections to $J_{M N}$ and Geodesics}\label{Jgen}

As has been already mentioned below (\ref{J1}), $J_{++}^{(1)}$ is conserved in the gravitational field of (\ref{s1}). In fact, conservation for this case happens to be valid to all orders in $\mu$ \footnote{In practice, only the first order in $\mu$ appears in the resulting equations. This is in accordance with our intuitive picture of figure \ref{vertex}: Gravity behaves linearly with respect to the metric (\ref{s1}).}
\begin{align}\label{con1}
\nabla^{M} J_{MN}^{(1)}=0.
\end{align}

\begin{figure}
\centering
\includegraphics[scale=0.7]{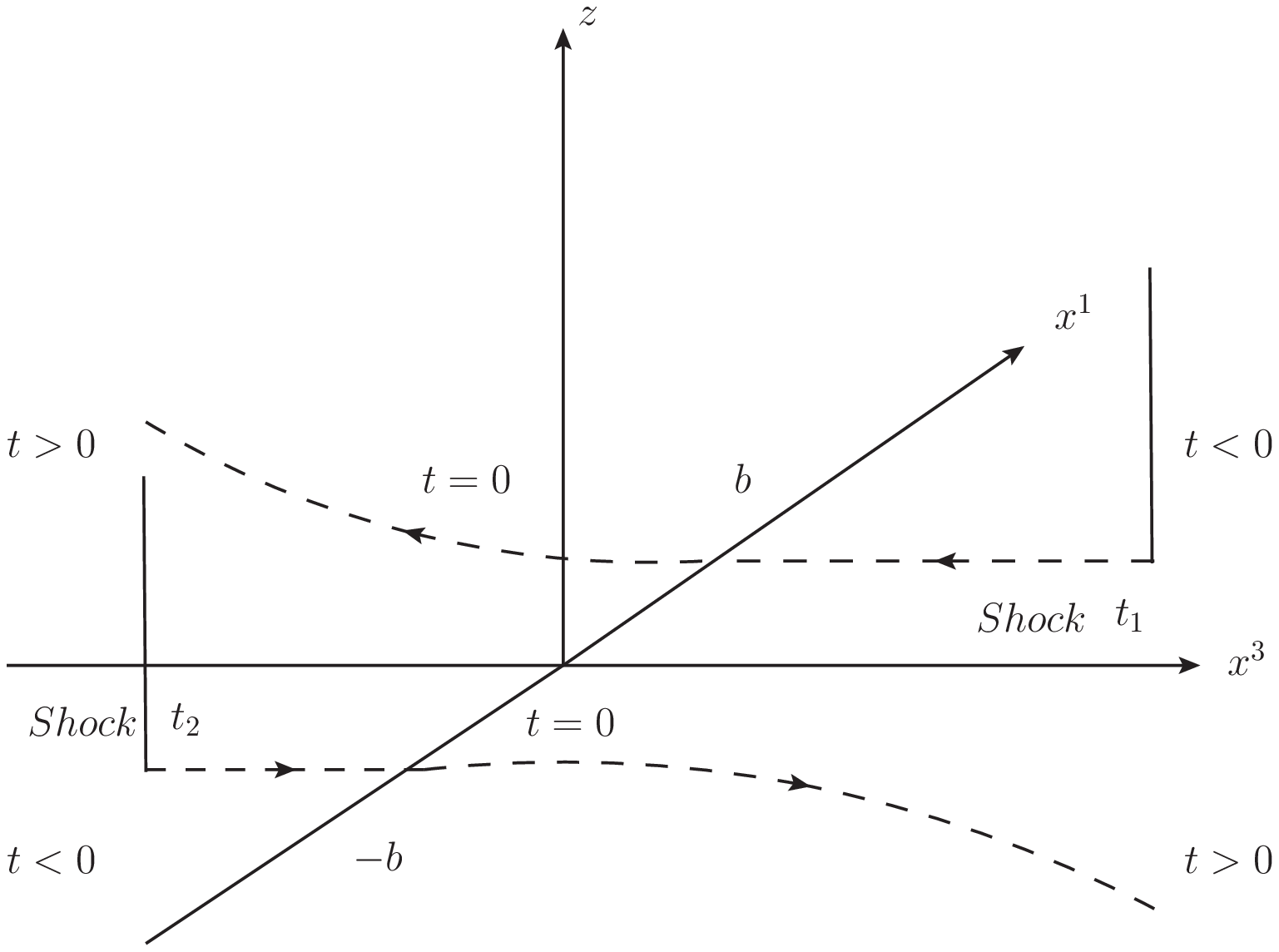}
\caption{The sources (one dimensional distributions of matter) represented as vertical lines extending along $z$ and their trajectories (suppressing the $x^2$-axis):  For negative times they move along straight (dotted) lines.  At t=0 each distribution intersects the shock due to the other distribution and its trajectory suffers a kick (see curved path); hence $J_{MN}$ changes with time.}
\label{BTen}
\end{figure}

\begin{figure}
\centering
\includegraphics[scale=0.8]{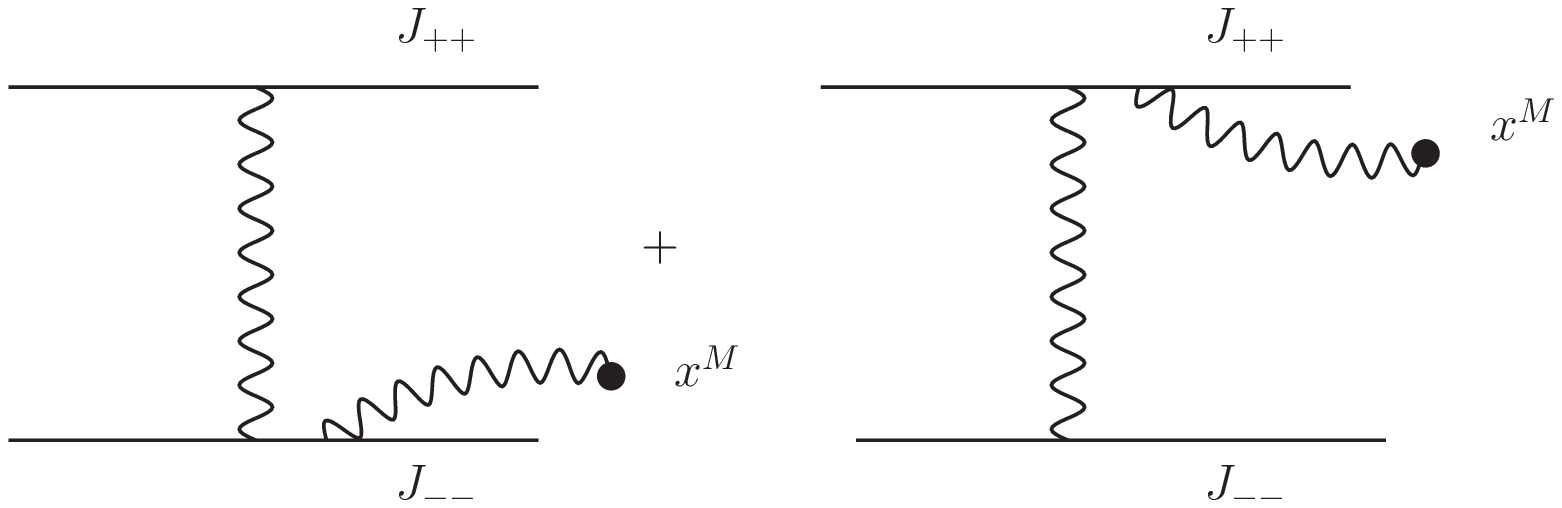}
\caption{Graviton emission diagrams resulting from self corrections to $J_{MN}$. The first source interacts via the gravitational field created by the other and vice versa. The point $x^M$ is the space-time point where $g_{M N}$ is measured. }
\label{Self_Int}
\end{figure}

Conservation to first order is still valid when we consider simultaneously the bulk distributions of matter $J_{++}^{(1)}$ and $J_{--}^{(1)}$ in the presence of the gravitational field (\ref{s12}). However, this is no longer true at the second order in $\mu$. The reason is because the $J_{--}$ ( $J_{++}$) source moves in the gravitational field of the $t_1$ $(t_2)$ , altering its initial trajectory. Figure \ref{BTen} outlines what happens while figure \ref{Self_Int} offers a diagrammatical intuition regarding the self-corrections to $J_{MN}$. This implies that we should correct $J_{MN}^{(1)}$ in order to preserve conservation (of the bulk Stress-Energy tensor). However, since we do not know the nature (equation of state) of $J_{MN}$ we make the assumption that these objects interact only via gravitational forces \footnote{More precisely, we assume that any other interactions are small compared to the gravitational forces.}.


\subsection{Calculating the Corrections for $J_{MN}$}\label{Tcor}
If the bulk matter we have to deal with consisted of just two point-like particles, they should have traveled along geodesics; as it is rigorously shown in \cite{Papa:1974} conservation of (the total) $J_{MN }$ is then guaranteed (for an application see \cite{Taliotis:2010az}). In this case we would need 

\begin{align}\label{Jppnoz}
J ^{MN} =m  \sum_{(I=1)}^2   \dot{x}_{(I)}^{M} \dot{x}_{(I)}^{N}  \frac{1} {\sqrt{-g}}\delta^{(4)} \left(\vec{x}_{(I)}- \vec{x}_{(I)}(s_{(I)}) \right) \hspace{0.2in}   x^{M}= (x^0,x^1,x^3,x^1,x^2,z)    
\end{align}
where $m$ has dimensions of mass while $\vec{x}$ includes the four spatial components. Equation (\ref{Jppnoz}) gives the total bulk {\bf SE} tensor of both point-like particles each one moving along the trajectory $ \vec{x}_{(I)}(s_{(I)})$ parameterized by $s_{(I)}$. The quantity $g$ is the determinant of the (total) metric tensor while the dots denote differentiation with respect to the parameter $s_I$. However, in our case we deal with one dimensional distributions of matter extending along the $z$ direction. Hence, we naturally generalize (\ref{Jppnoz}) to
\begin{align}\label{Jpp}
J ^{MN}= \frac{\pi \mu}{\kappa_5^2}  \sum_{(I=1)}^2   \dot{x}_{(I)}^{M} \dot{x}_{(I)}^{N}  \frac{1} {\sqrt{-g}}\delta^{(3)} \left(\vec{x}_{(I)}- \vec{x}_{(I)}(s_{(I)})\right)\rho^{MN}(z)  \hspace{0.2in}   \text{(no summing over M,N)}
\end{align}
where the factor $\frac{\pi \mu}{\kappa_5^2}$ has dimensions of mass, is first order in $\mu$ and its exact form is a convenient convention
 (see below) while $\vec{x}_{(I)}$ does not include the $z$ coordinate \footnote{In contrast with (\ref{Jppnoz}).}. In this equation we assume that the distribution of matter is made from point-like particles distributed in a continuous way in space-time along the $z$-axis. The total bulk {\bf SE} tensor is the summation (integral) of all of these particles which are assumed to be glued tight together so that there are no mutual interactions.
 
Under these assumptions, one may compute the $O(\mu^2)$ corrections to $J^{MN}$ from both distributions. Before doing so we find it instructive to check whether this formula reproduces (\ref{J1}) in the case of one distribution ($I=1$) which moves ultra-relativistically along negative $x^3$ and is parameterized by

 \begin{align}\label{tr1c}
x^{M}_{(1)}(x^0,x^3,x^1,x^2,z)=(t,-t,b,0,z).
\end{align}
In light cone coordinates and choosing to parameterize the trajectory by $x^-$, equation (\ref{tr1c}) then implies
 \begin{align}\label{tr1lc}
x^{M}_{(1)}=(x^+(x^-)=0,x^-(x^-),x^1(x^-)=b,x^2(x^-)=0,z(x^-))=(0,x^-,b,0,z).
\end{align}
Direct substitution of (\ref{tr1lc}) to (\ref{Jpp}) yields
\begin{align}\label{Tp1}
J ^{--}_{(1)}=J_{(1)++}= \frac{\pi \mu}{\kappa_5^2}z^4 \delta(x^+) \delta(x^1-b)\delta(x^2) 
\end{align}
for $\rho^{--}(z)=\int dz' \sqrt{-g(z')} z'^4 \delta(z'-z)=\sqrt{-g} z^4$ reproducing equation (\ref{J1}). This computation also clarifies the convention
 of $ \frac{\pi \mu}{\kappa_5^2}$ in (\ref{Jpp}). The second piece of $J_{\mu \nu}$, equation (\ref{J2}), due to the second particle is reproduced similarly.

The next step is to calculate the next (second) order corrections (in $\mu$) of $J_{\mu \nu}$ which translates into finding the corrections to the trajectories $x_{(I)}$. These may be obtained from the geodesic equations. In particular, we only need the first order corrections to $x_{(I)}$
as we already have a power of $\mu$ in front of the summation operator (see (\ref{Jpp})). The geodesic equations we need read

\begin{align}\label{geo}
\ddot{x^{M}}_I+\Gamma_{J;N P}^{M}  \dot{x_I^{N}}\dot{x_I^{P}}=0 \hspace{0.3in}I,J=1,2 \hspace{0.3in}I\neq J
\end{align}
and are interpreted as the motion of distribution $I$ in the gravitational field of the distribution $J$  (due to $\Gamma_{J;N P}^{M}$ where $\Gamma$ are the Christoffel symbols) and vice versa; this is precisely the meaning of the subscripts $I$ and $J$.
We begin with computing the corrections to the distribution $I=1$ whose (first order) perturbed trajectory looks like \footnote{We drop the subscript $_{(1)}$ which labels the particle $I=1$ for simplicity.}
 \begin{align}\label{tr11}
(x^{M}(x^-))^{(0)}+(x^{M}(x^-))^{(1)} =  \left((x^+)^{(1)},x^- + (x^-)^{(1)},b+(x^1)^{(1)},(x^2)^{(1)},(z)^{(1)} \right)
\end{align}
where we have chosen to parameterize the trajectory with $x^-$ (i.e. $s_{(I=1)}=x^-$). The superscript $^{(1)}$ denotes the order (in $\mu$) in the expansion. Taking into account (\ref{Jpp}) and the fact that $\Gamma_{(2)NP}^{M} \sim \mu$ otherwise is zero, we deduce that both of the terms $\dot{x^{N}}$ or $\dot{x^{P}}$ have to be of zeroth order; i.e. the only choice is $N=P=-$. This implies that we need to determine $\Gamma_{--}^{M}$ to first order in $\mu$ that arise from the second particle \footnote{Where we (also) dropped the subscript $_{(2)}$ which labels the second particle.} and which read
\begin{align}\label{Gam}
\Gamma_{--}^{+}=  -  \frac{z^4}{2} t_{2,x^-}     \hspace{0.3in}     \Gamma_{--}^{-}=  0  \hspace{0.3in}\Gamma_{--}^{1}=  -  \frac{z^4}{2} t_{2,x^1}   \hspace{0.3in}\Gamma_{--}^{2}= -   \frac{z^4}{2} t_{2,x^2}   \hspace{0.3in}\Gamma_{--}^{z}= -   z^3 t_2 .
\end{align}
A few explanations about our notation are in order: the term $t_2$ is due to the second particle and is given in equation (\ref{s12}). The subscript $_{2,x^{\mu}}$ on $t_{2,x^{\mu}}$ denotes ordinary differentiation of the source \footnote{As matter implies curvature and curvature implies matter, we will often use these notions interchangeably while their distinction should be evident from the context. } $t_2$ with respect to the coordinate $x^{\mu}$ . According to (\ref{s12}), the source $t_2$ is of first order in $\mu$ \footnote{From now on we drop the superscript $^{(1)}$ which denotes the order in $\mu$ from $t_{1,2}^{(1)}$.} and as a result, the same applies for the $\Gamma$'s of (\ref{Gam}); it should be by now obvious that the Christoffel symbols are due to the second distribution ($t_2$).

The final step is to integrate (\ref{geo}) using (\ref{Gam}) and using causal boundary conditions and substitute the result in formula (\ref{Jpp}). As we are interested in second order corrections in $\mu$, we immediately conclude that at least one of  $\dot{x^{M}}$ or $\dot{x^{N}}$ should be of order zero, i.e. $M=-$, $N \neq-$ or $M \neq-$, $N =-$. We also note that $\sqrt{-g}\sim O(1)+O(\mu^2)$ and hence according to (\ref{Jpp}) corrections from $\sqrt{-g}$ do not contribute to $J_{MN}$ at $O(\mu^2)$. We consider two cases

\vspace{0.3in}
\underline{Case I:  $\sqrt{-g}=O(1)$, $M=N=-$}
\vspace{0.3in}

In this case the modification of $J_{MN}$ of equation (\ref{Jpp}) to the order we are working is in the arguments of the delta's. We have  \footnote{ All the integrations with respect to $dx^{\pm}$ that follow from now on (see also Appendix \ref{A}) will imply the obvious: $\int dx^-$ stands for $\int_{-\infty}^{x^-} dx'^-$, $\int dx^- \int dx^-$ stands for $\int_{-\infty}^{x^-}dx'^-\int_{-\infty}^{x'^-} dx''^-$ etc.}
\begin{align}\label{JI1}
J ^{--}= &\frac{\pi \mu}{\kappa_5^2}  z^4  \delta \left(x^+ -   \frac{z^4}{2}\int dx^-  \int  dx^- t_{2,x^-} \right)              \delta \left(x^1-b -\frac{z^4}{2}  \int dx^-\int dx^- t_{2,x^1} \right)  \notag\\&           
 \hspace{0.7in} \times
 \delta \left(x^2 - \frac{z^4}{2}\int dx^-\int dx^- t_{2,x^2} \right)     + \text{the integrated $\delta(z)$ term}         
\end{align}
where by ``the integrated $\delta(z)$ term" we mean the contribution due to the deflection along $z$ integrated twice: the first integration takes into account that the point particle of the distribution $I=1$ located at $z$, interacts with the matter distribution $J=2$ and then the second integration sums over all particles of ``type" $I=1$. It turns out that this term is equal to $\frac{\pi \mu}{\kappa_5^2}z^6\int dx^- \int dx^- t_2\delta(x^+)\delta(x^1-b)\delta(x^2) $. Plugging this term in (\ref{JI1}) and expanding the delta's to first order in the sources we obtain
\begin{align}\label{JI2}
J ^{--}=& \frac{\pi \mu}{\kappa_5^2}z^4 \Bigg[ \delta(x^+)\delta(x^1-b)\delta(x^2)-\frac{z^4}{2}\int  dx^- \Bigg( t_{2} \hspace{0.02in} \delta'(x+)\delta(x^1-b)\delta(x^2) 
\notag\\&
+\int dx^- \Big( t_{2,x^1}\delta(x^+)\delta'(x^1-b)\delta(x^2)+ t_{2,x^2}\delta(x^+)\delta(x^1-b)\delta'(x^2) \Big)\Bigg)\notag\\&
 -8 z^2\int dx^- \int dx^- t_2  \delta(x^+)\delta(x^1-b)\delta(x^2)      \Bigg].
\end{align}
We may cast (the first order correction terms of the) last equation in a compact form by expressing it in terms of $t_1$ and $t_2$. Using the identity
\begin{align}\label{nab1}
\mu \delta(x^{\pm}) \delta^{(2)}(\vec{r} - \vec{b}_{1,2})=-\frac{1}{2 \pi} \nabla_{\bot}^2 t_{1,2}\hspace{0.3in}\vec{b}_{1,2}=(\pm b,0)
\end{align} 
(see (\ref{nablat}) and (\ref{s12})) the $O(\mu^2)$ terms of (\ref{JI2}) take the form

\begin{align}\label{JI3}
&(J_{(1)++})^{(2)}=(J_{(1)} ^{--})^{(2)}= \frac{1}{4 \kappa_5^2} \int  dx^- \notag\\&
 \times \left(z^8\left(  t_{2} \nabla_{\bot}^2 t_{1,x^+}  +\nabla_{\bot}^2 t_{1,x^1} \int dx^- t_{2,x^1}+\nabla_{\bot}^2 t_{1,x^2} \int dx^- t_{2,x^2}\right)+16z^6\nabla^2_{\bot}t_1\int dx^- t_2 \right)
\end{align}
where we restored the subscript $_{(1)}$ and the superscript $^{(2)}$ in order to highlight that this is the second order correction to $J_{MN}$ of the first distribution.

\vspace{0.3in}
\underline{Case II:  $\sqrt{-g}=O(1)$, $M=-$, $N \neq -$}
\vspace{0.3in}

In this case the modification of $J_{MN}$ of equation (\ref{Jpp}) to this order we are working is in the factor $\dot{x}^{M}\dot{x}^-=\dot{x}^{N}$. Combining (\ref{geo}) and (\ref{Gam}) one may compute $\dot{x}^{N}$. Plugging this result to (\ref{Jpp}) and employing the identity (\ref{nab1}) in order to write the transverse delta's in terms of $t_1$ yields
\begin{subequations}\label{JpII}
\begin{align}
&  (J^{+-}_{(1)})^{(2)}=  -   \frac{z^8}{4\kappa_5^2}t_{2}                  \nabla_{\bot}^2 t_{1} \label{JpII+-}\\&
    (J_{(1)+1})^{(2)}=-(J^{-1})^{(2)}=              \frac{z^8}{4 \kappa_5^2} \nabla_{\bot}^2 t_{1} \int  dx^- t_{2,x^1}        \label{JpII+1} \\ &
    (J_{(1)+2})^{(2)}=-(J^{-2})^{(2)} = 	        \frac{z^8}{4\kappa_5^2} \nabla_{\bot}^2 t_{1} \int  dx^- t_{2,x^2}. \label{JpII+2}\\&
  (J_{(1)+z})^{(2)}=-(J^{-z})^{(2)} = 	        \frac{z^7}{\kappa_5^2} \nabla_{\bot}^2 t_{1} \int  dx^- t_2. \label{JpII+z}
\end{align}
\end{subequations}

\vspace{0.1in}
\underline{Second order corrections to the total $J_{M N}$}
\vspace{0.3in}

The second order corrections to the stress energy tensor $(J_{(1)MN})^{(2)}$ of the first distribution is given by equations (\ref{JI3}) and (\ref{JpII}). The corrections $(J_{(2) MN})^{(2)}$ of the second distribution may be found analogously and therefore the second order corrections to the total $J_{MN}$ read
\begin{subequations}\label{Jmn2}
 \begin{align}
&  (J_{+-})^{(2)}= -  (J^{+-})^{(2)}=    \frac{1}{4}\frac{ 1}{\kappa_5^2}  z^8   \left(   t_{2} \nabla_{\bot}^2 t_{1} + t_{1} \nabla_{\bot}^2 t_{2} \right) \label{J+-}   \\&
 (J_{++})^{(2)}= \frac{1}{4\kappa_5^2} \int  dx^- \notag\\&
 \times \left(z^8\left(  t_{2} \nabla_{\bot}^2 t_{1,x^+}  +\nabla_{\bot}^2 t_{1,x^1} \int dx^- t_{2,x^1}+\nabla_{\bot}^2 t_{1,x^2} \int dx^- t_{2,x^2}\right)+16z^6\nabla^2_{\bot}t_1\int dx^- t_2 \right)
   \label{J++} \\&
  (J_{--})^{(2)}= \frac{1}{4\kappa_5^2} \int  dx^+ \notag\\&
 \times \left(z^8\left(  t_{1} \nabla_{\bot}^2 t_{2,x^-}  +\nabla_{\bot}^2 t_{2,x^2} \int dx^+ t_{1,x^2}+\nabla_{\bot}^2 t_{2,x^1} \int dx^+ t_{1,x^1}\right)+16z^6\nabla^2_{\bot}t_2\int dx^+ t_1 \right)      \label{J--}\\&
    (J_{+1})^{(2)}=              \frac{1}{4}\frac{1}{\kappa_5^2} \nabla_{\bot}^2 t_{1} \int  dx^- t_{2,x^1}
\hspace{0.45in}
   (J_{-1})^{(2)}=              \frac{1}{4}\frac{ 1}{\kappa_5^2} \nabla_{\bot}^2 t_{2} \int  dx^+ t_{1,x^1}         \label{J+-1}\\ &
    (J_{+2})^{(2)} = 	        \frac{1}{4}\frac{ 1}{\kappa_5^2} \nabla_{\bot}^2 t_{1} \int  dx^- t_{2,x^2}
\hspace{0.45in}
 (J_{-2})^{(2)} = 	        \frac{1}{4}\frac{ 1}{\kappa_5^2} \nabla_{\bot}^2 t_{2} \int  dx^+ t_{1,x^2}\label{J+-2}\\&
  (J_{+z})^{(2)} = \frac{z^7}{\kappa_5^2} \nabla_{\bot}^2 t_{1} \int  dx^- t_2 \hspace{0.45in}
  (J_{-z})^{(2)} = \frac{z^7}{\kappa_5^2} \nabla_{\bot}^2 t_{2} \int  dx^+ t_1  \label{J+-z}\\&
 (J_{11})^{(2)}= (J_{22})^{(2)}= (J_{12})^{(2)}= (J_{zz})^{(2)}=0
\end{align}
\end{subequations}
The first equality of (\ref{J+-}) is not completely obvious and so we prove it below by considering

\begin{align}\label{JTpm}
 (J_{+-})^{(2)}&=\left(g_{+M}g_{-N} J^{MN}\right)^{(2)} \notag\\&
=g_{+M}^{(0)}g_{-N}^{(1)} (J^{MN})^{(1)}+g_{+M}^{(1)}g_{-N}^{(0)} (T^{MN})^{(1)}+g_{+M}^{(0)}g_{-N}^{(0)} (J^{M N})^{(2)}\notag\\&
=g_{+-}^{(0)}g_{--}^{(1)} (J_{++})^{(1)}+g_{++}^{(1)}g_{-+}^{(0)} (J^{--})^{(1)}+g_{+-}^{(0)}g_{-+}^{(0)} (J^{-+})^{(2)}\notag\\&
=(-1)(z^4t_2) \left(-\frac{z^4}{2\kappa_5^2} \nabla_{\bot}^2t_1 \right)+(z^4t_2)(-1)\left(-\frac{z^4}{2\kappa_5^2} \nabla_{\bot}^2 t_1 \right)\notag\\&
+(-1)(-1) z^8\left( -\frac{1}{4}\frac{ 1}{ \kappa_5^2}     \left(   t_{2} \nabla_{\bot}^2 t_{1} + t_{2} \nabla_{\bot}^2 t_{1} \right) \right)  \notag\\&
= \frac{1}{4}\frac{ 1}{ \kappa_5^2}  z^8   \left(   t_{2} \nabla_{\bot}^2 t_{1} + t_{2} \nabla_{\bot}^2 t_{1} \right)
\end{align}
where in the fourth equality we used the fact that the (bulk) {\bf SE} tensor of the point particles (see (\ref{J1}), (\ref{J2}) and (\ref{nab1})) to first order in $\mu$ takes the form 
\begin{align}\label{J12nab}
J_{++}^{(1)}=-\frac{1}{2\kappa_5^2}z^4 \nabla_{\bot}^2 t_1 \hspace{0.3in} J_{--}^{(1)}=-\frac{1}{2\kappa_5^2}z^4 \nabla_{\bot}^2 t_2.
\end{align}
Although we will mostly work with (\ref{Jmn2}) which is a compact expression, for concreteness, we write $J_{MN}$ of (\ref{Jmn2}) in terms of the coordinates in order to clarify its form and show explicitly that it is localized (in all but the $z$ directions). Defining 
\begin{align}\label{r12}
 \vec{r_1}=   \vec{r}- \vec{b_1}               \hspace{0.3in}            \vec{r_2}=   \vec{r}- \vec{b_2} .
\end{align}
and employing (\ref{nab1}) in  (\ref{Jmn2}) we obtain

\begin{subequations}\label{Jmnc}
 \begin{align}
&  (J_{+-})^{(2)}=    \frac{\pi \mu^2}{2 \kappa_5^2}z^8\log(2k b) \delta(x^+)\delta(x^-)\left( \delta^{(2)}(\vec{r_1})  +   \delta^{(2)}(\vec{r_2})) \right), \label{J+-c}   \\&
(J_{++})^{(2)}=  \frac{\pi \mu^2}{2 \kappa_5^2} \theta(x^-)\Bigg[z^8 \log(2kb) \delta'(x^+)\delta^{(2)}(\vec{r_1}) +z^8 \frac{x^-}{x^1+b}\delta(x^+)\delta'(x^1-b)\delta(x^2) \notag\\&
\hspace{0.6in} +z^8 \frac{x^- x^2}{4b^2+(x^2)^2}\delta(x^+)\delta(x^1-b)\delta'(x^2) +16 x^- z^6 \log(2k|b|) \delta(x^+)\delta^{(2)}(\vec{r_1})    \Bigg] ,\label{J++c}\\&
    (J_{+1})^{(2)}= \frac{\pi \mu^2}{4 \kappa_5^2 b}z^8 \theta(x^-)\delta(x^+)\delta^{(2)}(\vec{r_1}),   \label{J+-1c}\\ &
    (J_{+2})^{(2)} = 	     0 ,\label{J+2c}\\&
 (J_{+z})^{(2)} = 	 \frac{2\pi \mu^2}{ \kappa_5^2}z^7 \theta(x^-)\log(2k|b|) \delta(x^+)\delta^{(2)}(\vec{r_1}) .  \label{J+zc}
%
\end{align}
\end{subequations}
The asymmetry between $J_{+1}^{(2)}$ and $J_{+2}^{(2)}$ is due to the fact that the impact parameter $\vec{b}$ has only $x^1$ component (see (\ref{s12})). The remaining nonzero components that complete (\ref{Jmnc}) may be obtained using the symmetries of the problem: $J_{--}^{(2)}$ and $J_{-1}^{(2)}$ may be obtained from $J_{++}^{(2)}$ and $J_{+1}^{(2)}$ respectively by interchanging $+ \leftrightarrow-$ and $b \leftrightarrow-b$. We  observe that $J_{MN}^{(2)}$ is localized in all but the $z$ direction and exists for positive times or exactly at the collision point at the collision time (this is the $J_{+-}^{(2)}$ component-see (\ref{J+-c})).



\subsection{Conservation, Tracelessness and Field Equations}

The second order corrections to (the total) $J_{MN}$ have already been calculated in the previous section. One may check that the action of a covariant derivative (contracted) on $J_{MN}$ yields

%
\begin{align}\label{Jcon}
\left((\nabla^{M})^{(0)}+ (\nabla^{M})^{(1)}\right) \left( (J_{MN})^{(1)}+ (J_{MN})^{(2)} \right) \sim \delta_{N \pm}\nabla_{\bot}^2 t_1 \nabla_{\bot}^2 t_2 +O(\mu^3)
\end{align}
where $\nabla$ denotes a covariant derivative, $\delta_{N \pm}$ is the Kronecker delta while the superscripts denote the order in $\mu$. According to (\ref{nab1}) the expression $\nabla_{\bot}^2 t_1 \nabla_{\bot}^2 t_2$ gives
\begin{align}\label{nab^2}
\nabla_{\bot}^2 t_1 \nabla_{\bot}^2 t_2 \sim \delta(\vec{r}-\vec{b}_1) \delta(\vec{r}-\vec{b}_2) \sim \delta(\vec{b}_2-\vec{b}_1) .
\end{align}
Therefore, conservation of $J_{M\pm}$ (to $O(\mu^2)$ at least) is valid if and only if the impact parameter is not zero. We thus see that introducing a nonzero impact parameter is essential in order to preserve conservation. This is one of the important conclusions that we derive in this work \footnote{Later we will see that a as $b \rightarrow 0$ we have to deal with another problem: the metric tensor also diverges.}.

It is also useful to compute the trace of $J_{MN}$ as it enters the field equations (see (\ref{Ein})). A short computation yields
\begin{align}\label{Tr}
J=g^{MN}J_{MN}=(g^{MN})^{(1)}(J_{MN})^{(1)}+(g^{MN})^{(0)}(J_{MN})^{(2)}=0+O(\mu^3)
\end{align}
which shows that the stress-energy tensor is traceless to order $\mu^2$. Tracelessness is very convenient as it simplifies Einstein's equations which become
\begin{align}\label{EE}
R_{MN}  = \kappa_5^2 J_{MN}+O(\mu^3)  \hspace{0.35in}\kappa_5^2=8\pi G_4.
\end{align}

\section{Field Equations}\label{FEq}

In this section we will write down the field equations (\ref{Ein}) to order $\mu^2$. These determine the functions $g_{\mu \nu}^{(2)}$ \footnote{$g_{\mu \nu}^{(2)}$ are not exactly a subset of the components of $g_{MN}$ since according to their defining equation (\ref{s12}), they are off by a factor of $\frac{z^2}{L^2}$.} of (\ref{s12}) which is what we need to calculate. We will write down their linearized form expanding around the background that is determined by the simultaneous presence of both shock waves and show how they may be decoupled. We will find out that  $g_{\mu \nu}^{(2)}$ obeys the equation

\begin{align}\label{box}
\Box g_{\mu \nu}^{(2)}=s_{\mu \nu}^{(2)}
\end{align}where $ \Box=\left(\frac{3}{z}\partial_{z}-\partial_{z}^2 - \nabla^2_{\bot} +2 \partial_{x^+}\partial_{x^-} \right)$ is the D'Alambertian in AdS$_5$ and $s_{\mu \nu}$ some tensor related to $J_{\mu \nu}$ of equation (\ref{Jmn2}). Although decoupling all of the metric components is possible and has been done, we only display the components of (\ref{EE}) that we will use. We need \footnote{From now on, we will drop the superscript (2) (which denotes the order in the expansion in $\mu$) on $g_{\mu \nu}^{(2)}$ and $J_{\mu \nu}^{(2)}$ for simplicity.}

\begin{subequations}\label{FE}
\begin{align}
(++)\hspace{0.15in} \frac{1}{2z}[ &3g_{++,z}-zg_{++,zz}-zg_{++,x^2x^2}-zg_{++,x^1x^1}\notag\\&
+zg_{+2,x^+ x^2}+zg_{+1,x^+x^1}-zg_{11,x^+x^+}-zg_{22,x^+x^+}]=\kappa_5^2J_{++},\label{++}\\
(+-)\hspace{0.15in} \frac{1}{4z}[ &-16z^z t_1t_2+ z^9 \left(t_{1,x^+}t_{2,x^-}-2t_{1,x^1}t_{2,x^1}-2t_{1,x^2}t_{2,x^2} \right)        \notag\\&
-2g_{11,z}  -2g_{22,z}+10g_{+-,z}-2zg_{+-,zz} -2zg_{+-,x^1x^1} -2zg_{+-,x^2x^2} \notag\\&
+zg_{+2,x^-x^2}+zg_{+1,x^-x^1}+zg_{-2,x^+x^2}+zg_{-1,x^+x^1}\notag\\&
-2zg_{++,x^-x^-}-2zg_{--,x^+x^+}+4zg_{+-,x^+x^-}-2zg_{11,x^+x^-}-2zg_{22,x^+x^-}]
=\kappa_5^2J_{+-},\label{+-}\\
(+1)\hspace{0.15in} \frac{1}{4z}[ &z^9t_{2,x^1}t_{1,x^+}+6g_{+1,z}-2zg_{+1,zz}-2zg_{+1,x^2x^2}+2zg_{+2,x^1x^2}
-2zg_{++,x^-x^1}\notag\\&+2zg_{12,x^+x^2}+2zg_{+-,x^+x^1}-2zg_{22,x^+x^1}+2zg_{+1,x^+x^-}-2zg_{-1,x^+x^+}]=\kappa_5^2J_{+1},\label{+1}\\
%
%
(11)\hspace{0.15in} \frac{1}{2z}[ &-8z^7 t_1t_2+z^9(t_{1,x^1}t_{2,x^1}+t_{1,x^1x^1}t_2+t_1t_{2,x^1x^1})\notag\\&
-2g_{+-,z}+4g_{11,z}+g_{22,z}-zg_{11,zz}-zg_{11,x^2x^2}+2zg_{12,x^1x^2}\notag\\&
+2g_{+-,x^1x^1}-zg_{22,x^1x^1}-2zg_{+1,x^-x^1}-2zg_{-1,x^+x^1}+2zg_{11,x^+x^-}]
=k_{5}^{2} J_{11}=0,\label{11}\\
%
%
%
(12)\hspace{0.15in} \frac{1}{4z}[ &z^9 \left( t_{2,x^2} t_{1,x^1}+ t_{1,x^2}t_{2,x^1}+2t_{2}t_{1,x^1x^2} +2t_{1}t_{2,x^1x^2} \right)
+6g_{12,z}-2zg_{12,zz}+4zg_{+-,x^1x^2}\notag\\&
 -2zg_{+1,x^-x^2}-2zg_{+2,x^-x^1}-2zg_{-1,x^+x^2}-2zg_{-2,x^+x^1}+4zg_{12,x^+x^-}]=\kappa_5^2J_{12}=0,
\label{12}\\
(zz)\hspace{0.15in}\frac{1}{2z}[&32z^7t_1t_2-2g_{+-,z}+g_{11,z}+g_{22,z}+2zg_{+-,zz}-zg_{11,zz}-zg_{22,zz}]=k_{5}^{2} J_{zz}=0,\label{zz}\\
(1z)\hspace{0.22in}\frac{1}{2}[&6z^7t_{1,x^1}t_2+6z^7t_{2,x^1}t_1+g_{12,x^2z}\notag\\&
+2g_{+-,x^1z}-g_{22,x^1z}-g_{+1,x^-z}-g_{-1,+z}]=k_{5}^{2} J_{1z},\label{1z}=0\\
(2z)\hspace{0.22in}\frac{1}{2}[&6z^7t_{1,x^2}t_2+6z^7t_{2,x^2}t_1+g_{12,x^1z}\notag\\&
+2g_{+-,x^2z}-g_{11,x^2z}-g_{+2,x^-z}-g_{-2,x^+z}]=k_{5}^{2} J_{2z}=0,\label{2z}\\
(+z)\hspace{0.22in}\frac{1}{2}[&2z^7t_{2}t_{1,x^+}+g_{+2,x^2z}+g_{+1,x^1z}-g_{++,x^-z}\notag\\&
+g_{+-,x^+z}-g_{11,x^+z}-g_{22,x^+z}]=\kappa_5^2J_{+z}\label{+z}
\end{align}
\end{subequations}
where the components of $J_{MN}$ are given by (\ref{Jmn2}). In order to decouple these equations we have to take appropriate linear combinations of derivatives or integrals (with respect to the coordinates $x^M$) of several of the components of (\ref{FE}). As a first step we integrate (\ref{zz}) twice with respect to $z$ from zero to z with the boundary condition that $g_{\mu \nu}$ vanishes as $z\rightarrow 0$. In particular, we apply the operator $z^2 \int_{0}^{z}1/z'^3 \int_{0}^{z'}2z'' dz''dz'$ in both sides of $(z''z'')$ of (\ref{zz}). The result is

\begin{align}\label{intz}
-2g_{+-}+g_{11}+g_{22}=\frac{2}{3}z^8 t_1t_2
\end{align}
This equation implies that the $z^4$ dependence of this equation is on the left hand side only; according to (\ref{hr}) this dependence is proportional to $T_{\mu}^{\mu}$. Therefore, equation (\ref{intz}) is the tracelessness condition of the $T_{\mu \nu}$ gauge theory tensor.

We now verify that conservation of $T_{\mu \nu}$ is compatible with (\ref{FE}). Let us consider for instance (\ref{1z}) and find the $z^4$ dependence of this equation which is given by
\begin{align}
\left(g_{12,x^2}+2g_{+-,x^1}-g_{22,x^1}-g_{+1,x^-}-g_{-1,x^+}\right)\big|_{z^4}=0 
\end{align}
and which together with the tracelessness condition ($-2g_{+-}+g_{11}+g_{22}\big|_{z^4}=0$) implies
\begin{align}\label{cons1}
\left(g_{12,x^2}+g_{11,x^1}-g_{+1,x^-}-g_{-1,x^+}\right)\big|_{z^4}=0. 
\end{align}
But equation (\ref{cons1}) is exactly the conservation equation for $T_{\mu \nu}$ at order $\mu^2$, that is
\begin{align}\label{cons2}
(\nabla^{\mu}T_{\mu \nu})^{(2)}=0  \hspace{0.3in} \nu=1.
\end{align}
In the same way one may see that $(\nabla^{\mu} T_{\mu \nu})^{(2)}=0$ for each $\nu=+,-,1,2$ and conservation is encoded in the  $(\nu z)$ components of the field equations and they hold order by order in $\mu$; consequently we have

\begin{align}\label{ec}
\nabla^{\mu}T_{\mu \nu}=0
\end{align}to all orders.

Next, we show how equations (\ref{FE}) may be decoupled. For this, we consider the linear combination $(11)-2\partial_{x^1}\left(\int_{0}^{z}(1z')dz'\right)$ \footnote{With this notation we mean that we integrate and differentiate (\ref{1z}) with respect to z and $x^1$ respectively and subtract the result from equation (\ref{11}). A similar meaning applies to all combinations that follow.}
which involves the components (\ref{11}) and (\ref{1z}) and employ equation (\ref{intz}) in order to get rid of the linear combination $2g_{12}+g_{22}$ that appears on the way. We apply a similar method for the $g_{22}$ component. For the $g_{12}$ we take $(12)-\partial_{x^2}\left(\int_{0}^{z}(1z')dz'\right)-\partial_{x^1}\left(\int_{0}^{z}(2z')dz'\right)$ where we make use of equations (\ref{12}), (\ref{1z}) and (\ref{2z}).
The $g_{+-}$ is decoupled by considering the combination $(+-)-\partial_{x^+}\left(\int_{0}^{z}(-z')dz'\right)-\partial_{x^-}\left(\int_{0}^{z}(+z')dz'\right)$ which makes use of (\ref{+-}), (\ref{+z}) and the corresponding equation for $(-z)$ \footnote {It may be obtained from $(+z)$ using the obvious discrete symmetries of the problem.}.
%
For the $g_{++}$ component, we take $(++)-2\partial_{x^+}\left(\int_{0}^{z}(+z')dz'\right)$ where we use equations (\ref{++}) and (\ref{+z}).
Finally for the component $g_{+1}$ we make use of the equations (\ref{+1}), (\ref{1z}) and (\ref{+z}) by considering $(+1)-\partial_{x^+}\left( \int(1z)dz\right)-\partial_{x^1}\left(\int(+z)dz\right)$ with a similar combination for the (+2) case. In all, cases, we make use of (\ref{intz}) in an appropriate way. After some algebra and using (\ref{Jmn2}) we eventually obtain
\footnote{The corresponding solutions for the components $g_{- \mu}$, $ \mu=-,1,2$ are obtained from $g_{+ \mu}$ by interchanging the coordinates $+\leftrightarrow-$ and setting the impact parameter $b \rightarrow-b$.}

\begin{subequations}\label{deq}
\begin{align}
\bigg(\frac{3}{z}\partial_{z}-\partial_{z}^2 &- \nabla_{\bot}^2 +2 \partial_{x^+}\partial_{x^-} \bigg)g_{11}=\frac{8}{3}z^6t_1t_2+ z^8 \left( \frac{2}{3}t_{1,x^1}t_{2,x^1}-\frac{1}{6}t_{1,x^1x^1}t_{2}-\frac{1}{6}t_{1}t_{2,x^1x^1}\right)   \label{x1}\\
\bigg(\frac{3}{z}\partial_{z}-\partial_{z}^2 &- \nabla_{\bot}^2 +2 \partial_{x^+}\partial_{x^-} \bigg)g_{12}=\frac{1}{3} z^8 \left(2t_{2,x^2}t_{1,x^1}+2t_{1,x^2}t_{2,x^1}-t_2t_{1,x^1x^2}-t_1t_{2,x^1x^2} \right)  \label{x12}\\
\bigg(  \frac{3}{z}\partial_{z}-\partial_{z}^2 &- \nabla_{\bot}^2 +2 \partial_{x^+}\partial_{x^-} \bigg)g_{++}=-\frac{1}{6}z^8t_2t_{1,x^+x^+} +8z^6  \nabla^2_{\perp} t_1 \int dx^- \int dx^-t_2  \notag\\&
\hspace{1.6in}+\frac{1}{2}z^8\int dx^- \int dx^- \left( t_{2,x^1}  \nabla^2_{\perp} t_{1,x^1} +  t_{2,x^2}  \nabla^2_{\perp} t_{1,x^2} \right)
 \label{x++}\\
\bigg(  \frac{3}{z}\partial_{z}-\partial_{z}^2 &- \nabla_{\bot}^2 +2 \partial_{x^+}\partial_{x^-} \bigg)g_{+1}=\frac{1}{3}z^8 \Big(-t_{2,x^1} t_{1,x^+} +2t_{2} t_{1,x^+x^1} \Big)\notag\\&
\hspace{1.6in}+\frac{1}{2}z^8\int dx^-  \left(   t_{2,x^1}  \nabla^2_{\perp} t_1 - t_2  \nabla^2_{\perp} t_{1,x^1}     \right)
 \label{x+1}\\
\bigg(  \frac{3}{z}\partial_{z}-\partial_{z}^2 &- \nabla_{\bot}^2 +2 \partial_{x^+}\partial_{x^-} \bigg)g_{+-}=-\frac{40}{3}t_1t_2z^6 +z^8\Big( \frac{2}{3}t_{1,x^+}t_{2,x^-}-t_{1,x^1}t_{2,x^1}-t_{1,x^2}t_{2,x^2}\notag\\&
\hspace{1.6in}-\frac{1}{4}t_1\nabla_{\bot}^2t_2 -\frac{1}{4}t_2\nabla_{\bot}^2t_1 \Big)
 \label{x+-}
\end{align}
\end{subequations}
The remaining differential equations (and hence solutions ) of $g_{\mu \nu}$ can be determined as follows: $g_{\mu 2}$ are obtained from $g_{\mu 1}$ by interchanging the coordinates $1 \leftrightarrow 2$ while the components $g_{- \mu}$ are obtained from components $g_{+ \mu}$ after interchanging simultaneously $x^+ \leftrightarrow x^-$ and $\vec{b}_1 \leftrightarrow \vec{b}_2$  where $\vec{b}_{1,2}$ are given by (\ref{nab1}). 
These equations will be solved in Appendices \ref{A} and \ref{C} \footnote{More precisely, we need just the $z^4$ coefficient of $g_{\mu \nu}$ and we will derive only this part of the metric.}.

\vspace{0.1in}
\underline{Remark:}
\vspace{0.1in}

The ($z^4$ coefficient of the) component $g_{+-}$ may and will be obtained from (\ref{intz}) and so we will not solve (\ref{x+-}). However, as a cross check we verify that (\ref{intz}) which has been obtained from (\ref{zz}), provides the same information as (\ref{x+-}). In order to see this, one may simply add equations (\ref{x1}) and the corresponding differential equation for $g_{22}$ and subtract twice (\ref{x+-}) yielding $\bigg(  \frac{3}{z}\partial_{z}-\partial_{z}^2 - \nabla_{\bot}^2 +2 \partial_{x^+}\partial_{x^-} \bigg)(g_{11}+g_{22}-2g_{+-}-\frac{2}{3}z^6 t_1t_1)=0$; this justifies the claim. In fact, this is a general feature of counting the degrees of freedom of both sides of the AdS/CFT duality correctly: there are 10 degrees of freedom for $T_{\mu \nu}$ in the gauge theory side while in principle in the gravity side there 15 degrees of freedom for the metric that are specified by 15 differential equations that result from the field equations. Hence there should be a redundancy. Obviously the redundancy for the metric comes from the coordinate choice freedom while 5 out of the 15 components of the field equations should be satisfied trivially. Indeed, in the Fefferman-Graham system we are using here, the trivial components of the field equations (\ref{EE}) are the $(zM)$ components of (\ref{deq}) and they encode, as already been said, the conservation and tracelessness of $T_{\mu \nu}$.

\section{Deriving $T_{\mu \nu}$ and results}\label{Cal}
\subsection{Deriving the Retarded Green's function}\label{RGF}
%
We wish to calculate the components $g_{\mu \nu}$ and of (\ref{deq}). In order to do this, we need to invert the differential operator $\left(\frac{3}{z}\partial_{z}-\partial_{z}^2 - \nabla^2_{\perp} +2 \partial_{x^+}\partial_{x^-} \right)$. In particular, we are looking for the Green's function of this operator with the boundary condition that it vanishes at $z=0$ and also at negative times $t$; that is we are looking for the retarded Green's function. This has already been done in \cite{ Danielsson:1998wt} \footnote{We would like to thank Robert Myers for bringing this to our attention.} and independently but later in \cite{Albacete:2009ji}. The integral representation of the retarded Green's function then is

\begin{align}\label{G1}
G (t,\vec{x}, z; t',\vec{x}', z') \, & = \theta (t - t') \, \frac{z^2}{z'} \,
\int\limits_0^\infty d m \int\frac{d^3k}{(2 \pi)^3} \notag \\ & \times \, m \frac{sin[(t-t') \sqrt{m^2+k^2}]}{\sqrt{m^2+k^2}}e^{i \vec{k} (\vec{x}-\vec{x}')} \, J_2 (m \, z)
\, J_2 (m \, z')
\end{align}
where $\vec{x}=(x^1,x^2,x^3)$. Performing the angular integration of the $k$ \footnote{It should not be confused with the IR cutoff of section \ref{trp}.} variable yields

\begin{align}\label{G2}
G (t,\vec{x}, z; t',\vec{x}', z') \, & = -\theta (t - t') \, \frac{z^2}{z'}\frac{1}{R}\partial_R
\int\limits_0^\infty d m \int_{-\infty}^{\infty} \frac{dk}{(2 \pi)^2} m \frac{sin[(t-t') \sqrt{m^2+k^2}]}{\sqrt{m^2+k^2}}\notag \\ & 
\times 
e^{i k R} \, J_2 (m \, z)
\, J_2 (m \, z'), \hspace{0.25in}R=|\vec{x}-\vec{x}'|.
\end{align}
Taking into account (\ref{nablat}) we find that $R$ from the previous equation satisfies

\begin{align}\label{Rr}
R^2=|\vec{r}-\vec{r'}|+(x^3-(x^3)')^2
\end{align}
where $x^1$ and $x^2$ are the coordinates on the transverse plane and $x^3$ is the collision axis. Finally, we perform the $k$ integration of (\ref{G2}). Working as in \cite{Albacete:2009ji} we obtain

\begin{align}\label{G3}
G (t,\vec{x}, z; t',\vec{x}', z') \, & = - \frac{1}{4\pi R} \,\theta (t-t') \partial_R \Big[\theta ((t-t')-R) \, \theta ((t - t')+R) \, \frac{z^2}{z'} \,
\int\limits_0^\infty d m \notag \\ & \times \, m \, J_0\left( m \sqrt{(t-t')^2-R^2} \right) \, J_2 (m \, z)
\, J_2 (m \, z')\Big].
\end{align}
It is easy to check that (\ref{G3}) is the correct Green's function using dimensional reduction: one may apply $2\pi \int_{0}^{\infty} dR R$ (integrate on the transverse plane) on $G$. The integral is trivial since it is a total derivative. The upper limit cancels because of the $\theta ((t-t')-R)$ and the final result is precisely the Green's function derived earlier in \cite{Albacete:2009ji}. Now, differentiating in (\ref{G3}) with respect to R we obtain

\begin{align}\label{G4}
G (t,\vec{x}, z; t',\vec{x}', z') \, & = \frac{1}{4\pi } \,\theta (t-t') \, \theta ((t - t')+R) \, \frac{z^2}{z'} \,
\int\limits_0^\infty d m \,m J_2 (m \, z) J_2 (m \, z')\notag \\ &
\times \left[- m J_1\left( m \sqrt{(t-t')^2-R^2} \right)\frac{\theta((t-t')-R)}{\sqrt{(t-t')^2-R^2}}+\frac{\delta((t-t')-R)}{R} \right]
\end{align}
where we have omitted one vanishing term. This Green's function shows explicitly that the alteration of the metric occurs for positive times and exists only in and on the forward light cone - it is causal as it should be! The second term on the right-hand side of (\ref{G4}) does not mix $x^{\mu}$ with $z$ directions and it is localized. This term may modify the part of the metric on the light-cone, that is it may modify the shock waves. However, as we will see, this term does not contribute to the {\bf SE} tensor of the gauge theory at the order $\mu^2$ that we are considering. Any (possible) modification of the shock waves occurs in the bulk and not on the boundary \footnote{The modification would not come with a $z^4$ dependence but with a higher power in z.}. Actually, this is one of our main conclusions for this paper and we will analyze it extensively in section \ref{VIC}.

Equation (\ref{G4}) provides the most useful form of the retarded Green's function for our purpose. Nevertheless, integration over $m$ is possible and so we perform the integral for completeness obtaining

\begin{align}\label{G5}
G (t,\vec{x}, z; t',\vec{x}', z') \, & = - \frac{1}{(2\pi)^2 R} \,\theta (t-t')\partial_R \Big[ \theta ((t-t')-R) \, \theta ((t - t')+R) \, \frac{z^2}{z'} \theta(s)\theta(2-s)\frac{1+2s (s-2)}{\sqrt{s(2-s)}}\Big] 
\end{align}with 

\begin{align}\label{s}
s=\frac{(t-t')^2-R^2-(z-z')^2}{2zz'}
\end{align}


The Green's function of (\ref{G4}) will be in general convoluted with powers of $z$. These integrals over $z$ may be performed and as we will see, the results involve hypergeometric functions that terminate when the power of $z$ is even. We have 

\begin{align}\label{zint}
\int _{0}^{\infty}z'^n G(z;z')dz'&= \frac{1}{4\pi } \,\theta (t-t') \, \theta ((t - t')+R) \notag \\ &
\times 
\left[\theta((t-t')-R)\Pi_n((t-t')^2-R^2,z^2) +z^n \frac{\delta((t-t')-R)}{R} \right]\notag\\&
= \frac{1}{4\pi } \Bigg[\theta\left( x^+-x'^+ \right) \theta\left( x^- -x'^- \right) \theta \left(\sqrt{2(x^+-x'^+)(x^--x'^-)}-|\vec{r}-\vec{r'}| \right) \notag\\&
\hspace{0.45in} \times \Pi_n(2(x^+-x'^+)(x^--x'^-)-|\vec{r}-\vec{r'}|^2,z^2) +z^n \frac{\delta((t-t')-R)}{R} \Bigg] \notag\\&
\hspace{2in}n=6,8,10 
\end{align}
where in the second equality we have used (\ref{Rr}) while we have defined
\begin{align}\label{P}
\Pi_n(u^2,z^2)=&(n-4) u^{-3+n/2}z^2\Bigg[ (u^2 -z^2) F\big(2-n/2,3-n/2,1,z^2/u^2\big)\notag \\ &
+\big(-u^2+(n-2)z^2\big) F\big(2-n/2,3-n/2,2,z^2/u^2\big) \Bigg ]\hspace{0.1in}n=6,8,10.
\end{align}
In equation (\ref{P}) $u^2$ is defined by
\begin{align}\label{tao}
u^2= 2(x^+-x'^+)(x^--x'^-)-|\vec{r}-\vec{r'}|^2 
\end{align}where $\vec{r}$ is given in (\ref{nablat}) while $F$ denotes hypergeometric functions. The above equation may be simplified in the case where $n$ is even and reduces to a simple polynomial which is not necessary to display explicitly here. Instead, we will restrict just to the $z^4$ coefficient since according to (\ref{GT1}) it encodes all the information (of the {\bf SE} tensor of the gauge theory) we want. Expanding in $z$ we obtain

\begin{align}\label{P4}
\Pi_n \big|_{z^4}=\frac{1}{8}n(n-2)(n-4) \left( 2(x^+-x'^+)(x^--x'^-)-|\vec{r}-\vec{r'}|^2 \right)^{-3+n/2}z^4
\end{align}
Indeed, we see that for even $n$, (\ref{P4}) is a simple polynomial of $x^{\mu}$. 
The second piece of (\ref{zint}) shows that it comes with a $z^n$ ($n>4$) dependence and so according to (\ref{GT1}) does not contribute (to the second order in $\mu$) to the gauge theory {\bf SE} tensor. However, it does modify the metric in the bulk and as a result it contributes to $T_{\mu \nu}$ at higher orders in $\mu$. Having performed the $z$ integration, we proceed to the rest of the coordinates. We organize the remaining integrations in the following two subsections but we will leave the details for the appendices.
%
\subsection{Integration over the Light-Cone Plane}\label{LCP}
%
We begin by introducing the following compact notation  
\begin{align}\label{Got}
G \otimes z'^n f(x'^{\mu}) \big|_{z^4}\equiv  \frac{1}{4 \pi} & \int_{-\infty}^{\infty}  dx'^+\int_{-\infty}^{\infty}dx'^- \int d^2 \vec{r}\hspace{0.01in}' \left(\Pi_{n}(x^{\kappa}-x'^{\kappa},z)  \big|_{z^4} \right)f(x'^{\mu})\notag\\&
\times \theta\left( x^+-x'^+ \right) \theta\left( x^- -x'^- \right) \theta \left(\sqrt{2(x^+-x'^+)(x^--x'^-)}-|\vec{r}-\vec{r'}| \right),
\end{align}
where $f(x^{\mu})$ is any arbitrary function of $x^{\mu}$ while the last integral denotes integration in the transverse plane. We wish to perform the $x^{\pm}$ integrations for all the possible cases that we will encounter while specifying $g_{\mu \nu}$ from (\ref{deq}). We organize these integrations in five cases while we leave the details of the calculation for Appendix \ref{A}.

%
%

\vspace{0.1in}
\underline{Remark 1:}
\vspace{0.1in}

The results of all of the integrations in the light-cone plane (performed in appendix \ref{A}) are proportional to the product $\theta(x^+)\theta(x^-)$. This implies that the second order corrections to $g_{\mu \nu}$ appear in the forward light-cone which is what we had initially demanded by seeking a causal solution (see (\ref{s12})).

\vspace{0.1in}
\underline{Remark 2:}
\vspace{0.1in}

The right-hand sides of (\ref{deq}) contains expressions of the form $t_1t_2$ differentiated with respect to $x^{\mu}$ in some fashion. According to (\ref{s12}), these expressions are proportional to $\delta(x^+)\delta(x^-)$ or their derivatives. Our previous analysis \footnote{Along with the work of Appendix (\ref{A}).} has already taken care of the $x^{\pm}$ integrations and so beginning from the following subsection, by $t_{1,2}$ we will mean just the transverse part of $t_{1,2}$: $-\mu \log(\sqrt{(x^1\pm b)^2+(x^2)^2})$.

\subsection{Integration over the Transverse Plane}\label{TP}

Having performed the $x^{\pm}$ integrations we move to the integration over the transverse plane. The quantities we have to integrate have the structure \footnote{There is another case where we have to integrate terms of the form $\nabla_{\bot}^2 t_{1,2}$. However, according to (\ref{nab1}) these (transverse) integrations are trivial as they involve delta functions.} $\partial_{x_a^i}(t_1t_2)$ or $\partial^2_{x_a^i x_c^j}(t_1t_2)$ where $a,c;i,j=1,2$. The subscript $a$ (c) and the superscript $i$ $(j)$ denotes differentiation of the source $t_a$ $(t_c)$ with respect to the space-time coordinate $x^i$ $(x^j)$. We may reduce the number of the different integrals that we have to perform by working as follows. We firstly introduce the vectors
\begin{align}\label{b}
\vec{b_1}=(b_{11},b_{12}) \hspace{0.3in}\vec{b_2}=(b_{21},b_{22}) 
\end{align}
and generalize the form of (the transverse part of) $t_{1,2}$ given by (\ref{s12}) to
\begin{align}\label{nt}
t_1(\vec{r}-\vec{b_1})=-\mu \log(k r_1)  \hspace{0.3in}  t_2(\vec{r}-\vec{b_2})=-\mu \log(k r_2)
\end{align}
where $\vec{r}_{1,2}$ were defined by (\ref{r12}). The next step is to exchange the derivatives acting on $t_{1,2}$, that is $\partial_{x_a^i}$ with differentiations with respect to $b$'s of (\ref{b}), that is with $-\partial_{b_{a i}}$\footnote{So for instance $t_{1,x^1}t_{2,x^1}$ takes the form $\partial^2_{b_{11} b_{21}}(t_1t_2)$.}. Finally, at the end of our calculations we take the limits
\begin{align}\label{lim}
 \vec{b_1} \rightarrow  (b,0) \hspace{0.3in}  \vec{b_2}\rightarrow  (-b,0). 
\end{align}
Looking at the right-hand side of equations (\ref{deq}) we see that they involve the product $t_1t_2$ differentiated with respect to the transverse coordinates \footnote{Where $t_1t_2 \sim \log(k r_1) \log(k r_2)$ (see (\ref{s12}) while the $x^{\pm}$ contributions have already been taken into account in the previous subsection (see Appendix \ref{A}).}. 
Exchanging the transverse differentiations, according to our earlier discussion in this subsection, with derivatives with respect to the components of $\vec{b}_{1,2}$ we see at once that we have to calculate the following integral
\begin{align}\label{Iin}
{\cal M}^{(n)} (r_1,r_2,\tau)=\frac{1}{2 \pi }\int_{0}^{\infty}\int_{0}^{2\pi}r'dr'd\phi' \theta(\tau-r')|(\tau^2 -r'^2)^n \log(k |\vec{r'}+\vec{r_1}|) \log(k |\vec{r'}+\vec{r_2}|)
\end{align}
with $n=0,2$ \footnote{The superscript $^{(n)}$ on ${\cal M}^{(n)}$ does not denote the order in the expansion in powers of $\mu$.} and where we have introduced the convenient factor $\frac{1}{2\pi }$.The integrations are performed in Appendix \ref{C} and the final result has the form
\begin{align} \label{M}
{\cal M}^{(n)} (r_1,r_2,\tau)& = \theta(r_1-\tau)\theta(r_2-\tau)\widetilde{{\cal M}}_1^{(n)} (r_1,r_2,\tau)+ \theta(\tau-r_2)\theta(r_1-\tau)\widetilde{{\cal M}}_2^{(n)} (r_1,r_2,\tau) \notag\\&
+ \theta(\tau-r_1)\theta(r_2-\tau) \widetilde{{\cal M}}_3^{(n)} (r_1,r_2,\tau)+\theta(\tau-r_1)\theta(\tau-r_1)\widetilde{{\cal M}}_4^{(n)} (r_1,r_2,\tau)
\end{align}
where the $\widetilde{{\cal M}}_i^{(n)}$'s may be found using (\ref{Ii}), (\ref{I}), (\ref{K}) and (\ref{Ki}). This is the last ingredient we need which allows us to obtain the desired solutions for equations (\ref{deq}). We display the results in the next section.


\vspace{0.5in}

\subsubsection*{The final formula for $T_{\mu \nu}$} 

\vspace{0.5in}

Calculating the {\bf SE} tensor is the main goal of our paper. The basic formula for our purpose is equation (\ref{hr}) whose right-hand side is specified using (\ref{deq}). The necessary equations we need are equations (\ref{nab1}), (\ref{Jmn2}), (\ref{r12}), (\ref{b}), (\ref{lim}), (\ref{c1}), (\ref{c2}), (\ref{c3}), (\ref{c4}), (\ref{c5}), (\ref{Ja}), (\ref{Ii}) along with (\ref{I}) and (\ref{K}) along with (\ref{Ki}). Since we are interested in the $z^4$ coefficient of $g_{\mu \nu}$ only, our starting point for specifying any component of $T_{\mu \nu}$ is equation (\ref{Got}). In particular for $T_{11}$ we use (\ref{c1a}) along with (\ref{Ii}) and (\ref{I}) and also (\ref{c1b}) along with (\ref{K}) and (\ref{Ki}). For $T_{12}$ we need (\ref{c1b}) along with (\ref{K}) and (\ref{Ki}). For the $T_{++}$ component we need (\ref{c4}) along with (\ref{Ja}), (\ref{c3b}) along with (\ref{nab1}) and also (\ref{c5}). Finally, for $T_{+1}$ we need (\ref{c2}) along with (\ref{Ii}) and (\ref{I}) and (\ref{c3a}) along with (\ref{nab1}). At the end of the integrations we should take the limits as in (\ref{lim}). The final result is

\begin{subequations}\label{SE}
\begin{align}
&\langle T_{11} ^{(2)} (x^{\mu}) \rangle =2\mu^2 \frac{N_c^2}{2\pi^2}\lim_{\vec{b}_{1,2}\rightarrow (\pm b,0)} \Bigg \{ 4 {\cal M}^{(0)}+\left[4\partial^2_{b_{11}b_{21}} -\partial^2_{b_{11}b_{11}}-\partial^2_{b_{21}b_{21}}\right] {\cal M}^{(2)} \Bigg \},
\label{T11}\\&
\langle T_{12} ^{(2)} (x^{\mu}) \rangle =4\mu^2 \frac{N_c^2}{2\pi^2}\theta(x^+)\theta(x^-)		\lim_{\vec{b}_{1,2}\rightarrow (\pm b,0)}  \Bigg \{  \left[2 \partial^2_{b_{11}b_{22}}+2 \partial^2_{b_{12}b_{21}} -\partial^2_{b_{11}b_{12}}-\partial^2_{b_{21}b_{22}}\right] {\cal M}^{(2)} \Bigg \},\label{T12}\\&
\langle T_{++} ^{(2)} (x^{\mu}) \rangle =2\mu^2 \frac{N_c^2}{2\pi^2}(x^-)^2\theta(x^+)\theta(x^-)		 \lim_{\vec{b}_{1,2}\rightarrow (\pm b,0)} \Bigg \{-2 {\cal J}(\tau) 
+3\frac{(\tau^2-r_1^2)^2}{\tau^4} \theta(\tau - r_1)\notag\\&
\hspace{2.3in} \times \left[2-\frac{(b_{11}-b_{21})(x^1-b_{11})+(b_{12}-b_{22})(x^2-b_{12})}{|\vec{b}_1-\vec{b}_2|^2}\right] \Bigg \},\label{T++}\\&
\langle T_{+1} ^{(2)} (x^{\mu}) \rangle =2\mu^2 \frac{N_c^2}{2\pi^2}x^- \theta(x^+)\theta(x^-)		 \lim_{\vec{b}_{1,2}\rightarrow (\pm b,0)} \Bigg \{4 \left(\partial_{b_{21}}-2\partial_{b_{11}}\right) {\cal M}^{(0)} + 3 \frac{r_1^2-\tau^2}{\tau^2}\theta(\tau-r_1) \notag\\& \hspace{2in}
\times \left[ (b_{11}-b_{21}) \frac{r_1^2-\tau^2}{|\vec{b}_1-\vec{b}_2|^2} +
2(b_{11}-x^1)\log\left(k|\vec{b}_1-\vec{b}_2|\right) \right]     \Bigg \},\label{T+1}\\&
\hspace{2in}  \langle T_{+-} ^{(2)} (x^{\mu}) \rangle =\frac{1}{2}\left(\langle T_{11} ^{(2)} (x^{\mu}) \rangle +\langle T_{22} ^{(2)} (x^{\mu}) \rangle\right)
\end{align}
\end{subequations}
where ${\cal J}={\cal J}(\tau,r_1,r_2)$, ${\cal M}^{(0)}={\cal M}^{(0)}(\tau,r_1,r_2)$ and ${\cal M}^{(2)}={\cal M}^{(2)}(\tau,r_1,r_2)$ are given by (\ref{Ja}), (\ref{I}) and (\ref{K}) respectively while the equation for $T_{+-}^{(2)}$ follows from the tracelessness condition, equation (\ref{intz}). The components $T_{\mu2}^{(2)}$ are obtained from $T_{\mu1}^{(2)}$ when interchanging the labels $1 \leftrightarrow 2$ while components $T_{\mu-}^{(2)}$ are obtained from $T_{\mu +}^{(2)}$ when interchanging simultaneously the $+,\vec{b}_1 \leftrightarrow -,\vec{b}_2$ before taking the limits $\vec{b}_{1,2}\rightarrow (\pm b,0)$. Formula (\ref{SE}) is our final result which we will analyze in the next section.
\section{Discussion}\label{VIC}

\subsection{Regime of Validity}\label{va}

In section \ref{trp} we argued that the {\bf SE} tensor of the single nucleus can be trusted for transverse distances $r<\frac{1}{k}$. Therefore, in the case where we have two nuclei colliding with an impact parameter $b$, we trust the solution for

\begin{align}\label{k}
\tau,\hspace{0.03in}r_{1,2}\ll \frac{1}{k} \hspace{1in} r_{1,2}=\sqrt{(x^1\mp b)+(x^2)^2}.
\end{align}

On the other hand, from dimensional analysis, we expect that higher powers in $\mu$ should be compensated by higher powers in $\tau$, $r_1$ and $r_2$ which  generally will be multiplied by logarithms with arguments $\frac{r_1}{r_2}$,$\frac{r_{1,2}}{\tau}$  $k r_{1,2}$, $kb$ and $k \tau$ which we will collectively call $q$. In particular, we expect that the $\bf{SE}$ tensor at mid-rapidity should behave like

\begin{align}\label{T}
T_{\mu \nu} = \mu^2 \sum_{kl m np} c^{(2)}_{klmnp} \tau^kb^l r_1^m r_2^n \log^p[q^{(2)}_{klmnp}] + \mu^3 \sum_{kl m np} c^{(3)}_{klmnp} \tau^k b^l r_1^m r_2^n \log^p[q^{(3)}_{klmnp}] + \ldots
\end{align}
where $k$, $l$, $m$, $n$ are integers and $c_{klmnp}$ real numbers. In each sum the superscript $( j)$ where $ j=2,3,\ldots$ denotes the order in the expansion in powers of $\mu$. From dimensional analysis, at any given order in $\mu$ \footnote{Recall that $\mu$ has dimension of mass cubed.}, the powers should add up to $3j-4$, i.e. $k+l+m+n=3j-4$. In addition, in order to have a finite energy production at any finite piece of the space-time, we expect that when a negative power of one of the variables $\tau$, $r_1$ or $r_2$ appears, it should be compensated by a different variable with positive power that tends to zero faster; the term in the square bracket of equation (\ref{T++}) provides such an example. Mathematically this is achieved by the presence of theta functions. As a result, this fact shows how causality and conservation conspire in order to prevent the produced $T_{\mu \nu}$ from becoming infinite. However, this does not mean that $r_1$ and $r_2$ cannot be larger that $\tau$. What it is implied here is that when $r_{1,2}>\tau$ then $r_{1,2}$ will appear as denominators while $\tau$ as a numerator; the term $\left( \theta(r_2-r_1)\theta(r_1-\tau) +\theta(r_1-r_2)\theta(r_2-\tau) \right ) {\cal M}^{(0)}_1(\tau)$ of equation (\ref{I}) with ${\cal M}^{(0)}_1({\tau})$ given by (\ref{I1}) is an example of this case.

In analogy to \cite{Albacete:2008vs} and \cite{Albacete:2009ji}, we may imagine the gravitational description of the process as a shower of graviton exchanges between the sources and the bulk where the gravitational field is measured.  Below, we will argue that an expansion of  the form of equation (\ref{T}) is consistent with our diagrammatical intuition. The first term of the right-hand side of equation (\ref{T}) corresponds to figure \ref{interaction}. At central rapidities the effective vertices (see figure \ref{vertex}), loosely speaking, should have as dimensionless couplings the quantities

\begin{align}\label{coup}
\mu \tau^3 | \log(k r_1)| \hspace{0.3in} \mu \tau^3 | \log(k r_2)|.
\end{align}

We firstly note that the couplings of (\ref{coup}) being small is equivalent to

\begin{align}\label{coup1}
\mu \tau^3 | \log(k r_1)| \ll1 \hspace{0.3in} \mu \tau^3 | \log(k r_2)| \ll1 \hspace{0.3in}\mu \tau^3 | \log(k r)|\ll 1 \hspace{0.3in} \mu \tau^3 | \log(k b)| \ll 1.
\end{align}
We stress that the last two terms of the previous equation are a consequence of the first two and are not extra conditions. Now, because of (\ref{k}), the absolute value of the logarithms is greater than unity and as a result (\ref{coup1}) holds provided that

\begin{align}\label{mt}
\mu \tau^3 \ll 1.
\end{align}
Equation (\ref{mt}) justifies most of the terms in the expansion of (\ref{T}). The negative powers of $r_{1,2}$ and $\tau$ exist by dimensional analysis while their finite contribution to $T_{\mu \nu}$ has been already justified by causality and conservation.

We still have to investigate the conditions under which the inverse powers of $b$  (see (\ref{T++}) and (\ref{T+1})) do not cause the breakdown of the perturbative expansion (in $\mu$). It seems that the inverse powers of $b$ appear when the logarithms are differentiated with respect to $x^1$ or $x^2$ and are compensated by $x^{1,2}\theta(\tau-r_{1,2})$ or $\tau-r_{1,2}\theta(\tau-r_{1,2})$. Thus that as long as

\begin{align}\label{bt}
\mu \tau^3 \frac{\tau}{b} \ll 1
\end{align}
we may trust the first terms of ({\ref T}). On the other hand, unless we calculate the next order correction of $T_{\mu \nu}$ in $\mu$, we cannot be certain whether the inverse power of $b$ appearing in (\ref{T++}) and (\ref{T+1}) iterates or whether is a harmless overall factor appearing just once at order $\mu^2$. In the latter case, restriction (\ref{bt}) is not required.

As $b \rightarrow 0$ the metric exhibits in addition a logarithmic divergence. From a physics point of view, a similar logarithmic divergence is observed in gluon exchanges between two sources (such as quarks) in the eikonal approximation as the two sources approach each other \cite{Kovchegov:1997ke}. Both types of divergences are due to the corrections of the bulk {\bf SE}, $J_{MN}$.

It is also interesting to consider the high energy limit. Taking into account (\ref{kL}) and (\ref{mu}) we have that

\begin{align}\label{mk}
\mu^{\frac{1}{3}} \gg k.
\end{align}
We note that the inequalities (\ref{mt}) and (\ref{k}) are consistent with (\ref{mk}) and when combined yield to

\begin{align}\label{kmt}
k \tau \ll \mu^{\frac{1}{3}}\tau \ll 1.
\end{align}

What one has to keep from the above discussion is that the first terms of $T_{\mu \nu}$ in (\ref{T}) hold for sufficiently small proper times and high energies (compared to the transverse scale $k$) such that inequalities (\ref{k}), (\ref{coup1}), (\ref{bt}) and (\ref{mk}) apply simultaneously  (equivalently when (\ref{kmt}) applies). In particular, (\ref{k}) justifies the choice of the transverse profile of the  and (\ref{mk}) the high energy limit. Finally, (\ref{coup1}) associates the linear approximation of the gravitational field with a small proper time expansion. This makes physical sense since the gravitational field in the linear approximation is initially good enough to describe the process until the nonlinearities of the field will dominate for later times; this is just like the McLerran-Venogopolan model \cite{McLerran:1993ka} when analyzed at early times \cite{Lappi:2003bi}.

\subsection{Investigating the solution }\label{IS}

 The energy density  $\epsilon$ of the produced medium may be found by the formula

\begin{align}\label{ed}
\epsilon \equiv T_{00}= \frac{1}{2} \left(T_{++}+2 T_{+-} +T_{--}\right)= \frac{1}{2} \left(T_{++}+ T_{11}+ T_{22}+ T_{--}\right)
\end{align}
where the second equality is the tracelessness condition (\ref{intz}). Unfortunately a complete analysis of the results appears to be rather involved and is left for a separate project. Nevertheless there are some kinematical regions in which we can derive some conclusions. In particular, we investigate the energy density $\epsilon$ for the kinemtical regions $\tau \ll r_1, r_2$ and $r_1,r_2 \ll \tau$. According to figure \ref{re} these two regions belong to the regions $I$ ($I$$^{'}$) and $III$ ($III$$^{'}$) respectively \footnote{The reason they come in pairs is because there is an obvious $r_1 \leftrightarrow r_2$ mirror symmetry.}.  Certainly in either case, (\ref{k}), (\ref{coup1}), (\ref{bt}) and (\ref{mk}) are assumed to be still applicable.

 \FIGURE{\includegraphics[width=12cm]{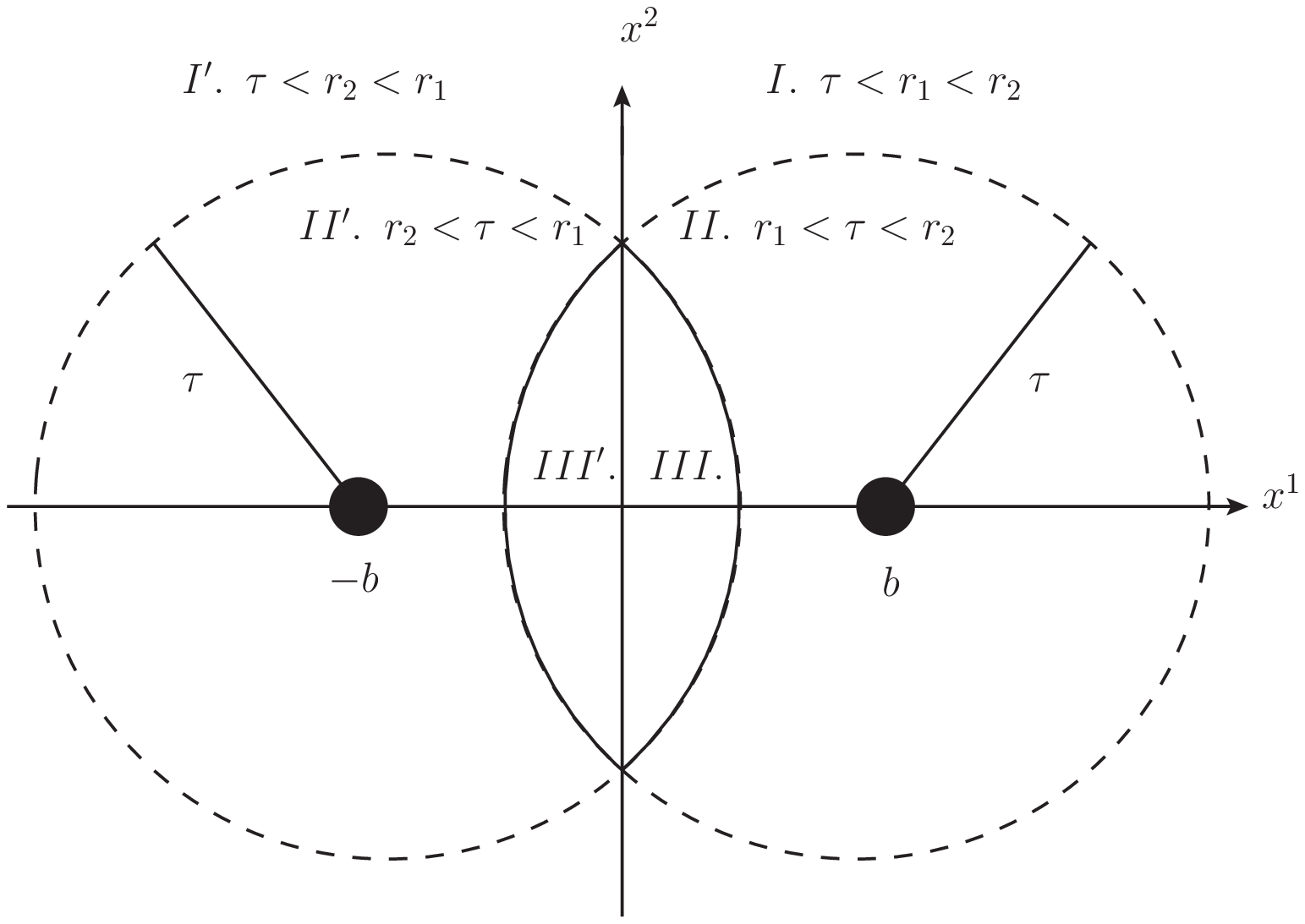}
  \caption{The reaction plane: Region $III$ corresponds to $r_1 < r_2 < \tau$ while region $III$$'$ corresponds to $r_2 < r_1 < \tau$. The dark dots are the centers of the nuclei at impact parameter 2b while ${r}_{1,2}$ denote the distance of the arbitrary point $\vec{r}$ from the center of each nucleus (right and left respectively). At any given proper time $\tau$, the propagation from the centers will reach the points on the peripheries (at most). This suggests that any given point $\vec{r}$ on the transverse plane of the produced medium at given $\tau$ will evolve according to the region to which it belongs. We find that our calculations are consistent with this expectation (see (\ref{TMN})).
}
\label{re}}

We argue that the leading contributions to $T_{\mu \nu}$ in the two regions that we will investigate, come from the quantities ${\cal J}$ and ${\cal M}^{(0)}$ which arise as a consequence of the initial shock waves $t_{1,2}$ and not because of $J_{MN}$ or ${\cal M}^{(2)}$. The reason is because these quantities come with a double (large) logarithmic enhancement in the kinematical areas under consideration. The rest of the terms in $T_{\mu \nu}$ either vanish exactly because of the presence of $\theta$ functions or do not exhibit this enhancement. %
Table \ref{Il} exhibits the double logarithmic behavior of ${\cal M}^{(0)}$. An analogous enhancement exists for the factor ${\cal J}$ given by equation (\ref{J}). The logarithmic suppression of the quantity ${\cal M}^{(2)}$, equation (\ref{K}), is not evident.  Thus, we have constructed table \ref{Kl} and table \ref{IK}, in appendix \ref{C} in order to clarify the behavior of ${\cal M}^{(2)}$. One has to keep in mind that when ${\cal M}^{(2)}$ appears, it has to be differentiated twice as shown in (\ref{T11}) and (\ref{T12}) and then appropriate limits are taken as in (\ref{lim}). This procedure lowers the logarithmic enhancement of these terms so in this approximation they are ignored.

\subsubsection {Energy density for regions $I$ and $I$$^{'}$}

We are interested in
 the energy density at the regions defined by \footnote{Equation (\ref{k}) still applies.} 

\begin{align}\label{et}
r_{1} \gg \tau \hspace{0.3in} r_2 \gg \tau.
\end{align}
According to figure \ref{re}, we are looking at (a sub-region of) regions $I$ and $I$$^{'}$. Mathematically this means that we are interested in
 the terms that are multiplied by $\theta(r_{1,2}-\tau)$.

\underline{$T_{11}$ and $T_{22}$ contributions:}
\vspace{0.15in}

The largest contribution to $T_{11}$ (and $T_{22}$), according to our previous discussion and equation (\ref{T11}), comes from the coefficient ${\cal M}^{(0)}$. Using table \ref{Il} we find

\begin{align}\label{ettr}
T_{11}=T_{22}&=2\mu^2 \frac{N_c^2}{ \pi^2}\Big \{ \left( \theta(r_2-r_1) \theta(r_1-\tau)+\theta(r_1-r_2) \theta(r_2-\tau)  \right)\tau^2  \log(kr_1)\log(kr_2)\notag\\&
+O(\mu^2\frac{\tau^4 }{r_1 r_2})\Big \}.
\end{align}

\underline{$T_{++}$ and $T_{--}$ contributions:}
\vspace{0.15in}

In this case the term that gives the largest contribution to $T_{++}$  is the term proportional to ${\cal J}$ \footnote{Similarly for $T_{--}$ whose main contribution is obtained from that of $T_{++}$ by $-b,\leftrightarrow b$ and $+ \leftrightarrow -$.} of appendix \ref{C}. This term contains the factor $\log(kr_1)\log(kr_2)$ and hence we deduce that

\begin{align}\label{ettpm}
T_{\pm \pm}&=-2\mu^2 \frac{N_c^2}{ \pi^2} (x_{\pm})^2 \Big\{ \left( \theta(r_2-r_1) \theta(r_1-\tau)+\theta(r_1-r_2) \theta(r_2-\tau)  \right)  \log(kr_1)\log(kr_2)\notag\\&
+O(\mu^2 \frac{\tau^2 }{r_1 r_2})\Big \}.
\end{align}
The last step is to substitute (\ref{ettr}) and  (\ref{ettpm})  in (\ref{ed}) to find that

\begin{align}\label{eted}
\epsilon_{I}=\epsilon_{I'} &=2\mu^2 \frac{N_c^2}{ \pi^2}  (2\tau^2-(x_+)^2-(x_-)^2) \Big \{ \left( \theta(r_2-r_1) \theta(r_1-\tau)+\theta(r_1-r_2) \theta(r_2-\tau)  \right)  \log(kr_1)\log(kr_2)\notag\\&
+O(\mu^2\frac{\tau^4 }{r_1 r_2}) \Big \}.
\end{align}
We observe that the energy density is symmetric under $r_1 \leftrightarrow r_2$ in regions $I$ and $I$$^{'}$. The diagonal elements of $T_{\mu \nu}$ in the $x^0,x^1,x^2,x^3$ coordinates at central rapidities where $(x^+)^2 \sim (x^-)^2 =\frac{1}{2}\tau^2$ obey

\begin{align}\label{all=}
 T_{00}=T_{11}=T_{22}=-3T_{33} \Big|_{x_+=\hspace{0.02in}x_-}
\end{align}
and receive their main contribution from the center of each nucleus separately in a multiplicative way as equations (\ref{ettr}) and (\ref{ettpm}) show.

\subsubsection {Energy density for regions $III$ and $III$$^{'}$}
Analogously to the previous case we are interested in
 the energy density at the regions defined by\footnote{Equation (\ref{k}) still applies.} 

\begin{align}\label{lt}
r_{1} \ll \tau \hspace{0.3in} r_2 \ll \tau
\end{align}
which implies 

\begin{align}\label{x12b}
x^1,\hspace{0.03in}x^2, \hspace{0.03in}b \ll \tau .
\end{align}
Inequality (\ref{lt}) implies that the propagation from the center of both nuclei has enough proper time $\tau$ in order to travel and affect the arbitrary point $\vec{r}$ on the transverse plane. The inequality (\ref{x12b}) shows that for the case under consideration, we may derive results for the area around the origin, that is around the center of the collision. This kinematical area concerns regions $III$ and $III$$^{'}$ of figure \ref{re}. Mathematically this means that the contributing terms should be multiplied by $\theta(\tau-r_{1,2})$.

\underline{$T_{11}$ and $T_{22}$ contributions:}
\vspace{0.15in}

Working as we did for regions $I$ and $I$$^{'}$, we find that

\begin{align}\label{lttr}
T_{11}=T_{22}&=2\mu^2 \frac{N_c^2}{ \pi^2} \Big\{ \left( \theta(\tau-r_2) \theta(r_2-r_1) +\theta(\tau-r_1)\theta(r_1-r_2)  \right)\tau^2  \log(k \tau)\log(k \tau)\notag\\&
+O\left(\mu^2\tau^2  \log(k r_{1,2})\right) \Big \}
\end{align}
where we have assumed that $|\log(kr_{1,2})| \sim |\log(k b)|\ll |\log^2(k \tau)|$ although in this region $|\log(kr_{1,2})| \sim |\log(k b)|> | \log(k \tau)|$.

\underline{$T_{++}$ and $T_{--}$ contributions:}
\vspace{0.15in}

The procedure is the same as before. Employing equation (\ref{J}) we have 

\begin{align}\label{ltpm}
T_{\pm \pm}&=-\mu^2 \frac{N_c^2}{ \pi^2} (x_{\pm})^2\Big\{ \left( \theta(\tau-r_2) \theta(r_2-r_1) +\theta(\tau-r_1)\theta(r_1-r_2)  \right)  \log(k \tau)\log(k \tau)\notag\\&
+O(\mu^2\frac{\tau^2} {r_1 r_2}) \Big \}.
\end{align}
Finally, substituting (\ref{ltpm}) and (\ref{lttr}) in equation (\ref{ed}) we deduce that

\begin{align}\label{lted}
\epsilon_{III}=\epsilon_{III'}&=\mu^2 \frac{N_c^2}{ \pi^2}(2\tau^2-(x_+)^2-(x_-)^2)  \Big\{ \left( \theta(\tau-r_2) \theta(r_2-r_1) +\theta(\tau-r_1)\theta(r_1-r_2)  \right)  \log(k \tau)\log(k \tau)\notag\\&
+O(\mu^2 \tau^2 \log(kr_{1,2})) \Big \}.
\end{align}

The conclusions are similar as in the previous case. The difference is that the arguments of the logarithms (nuclear profiles) have been changed to proper time $\tau$. This means that there is enough proper time $\tau$ for the propagation from the center of each nucleus to the observation point. At central rapidities we again have that

\begin{align}
 T_{00}=T_{11}=T_{22}=-3T_{33}\Big|_{x_+=\hspace{0.02in}x_-}.
\end{align}

\vspace{1in}

\subsection{Momentum anisotropy and spatial eccentricity}\label{Eccsec}


The momentum anisotropy $\epsilon_{\rho}(\tau)$ and spatial eccentricity $ \epsilon_{x}(\tau)$ of the produced medium are defined by
\begin{align}\label{man}
\epsilon_{\rho}(\tau)=\Bigg|\frac{\int_{-\infty}^{\infty}dx^1dx^2(T_{11}-T_{22}) }{\int_{-\infty}^{\infty}dx^1dx^2(T_{11}+T_{22})}\Bigg|,
\hspace{0.3in}
\epsilon_{x}(\tau)=\frac{\int_{-\infty}^{\infty}dx^1dx^2 T_{00}((x^1)^2-(x^2)^2) }{\int_{-\infty}^{\infty}dx^1dx^2 T_{00}((x^1)^2+(x^2)^2)}.
\end{align}
Having found the formula of $T_{\mu \nu}$, we are in the position to plot both of these quantities as a function of proper time $\tau$ and the impact parameter $b$. As the underlying expressions are complicated, we leave this for a future paper. For the moment we restrict ourselves to finding the momentum anisotropy and spatial eccentricity in the two extreme cases discussed in the previous subsection. 

The equality $T_{11}=T_{22}$ in both of these cases applies very close to the collision center or very far from it. Both are considered at sufficiently early times (compared to the energy of the collision/see subsection \ref{va}). Formula (\ref{man}) shows that that the momentum anisotropy in either case is zero, that is
\begin{align}\label{MA}
\epsilon_{\rho}(\tau)\Big|_{b \gg \tau}\approx 0 \hspace{0.03in}, \hspace{0.3in} \epsilon_{\rho}(\tau)\Big|_{b \ll \tau}  \approx 0.
\end{align}
The physical meaning of this result is that at the early stages of the collision (where our analysis is applicable) and at the center of the collision, the system behaves isotropically until it will starts to feel the anisotropy from the edges of the nuclei at later times. This result agrees with \cite{ Kolb:2000sd }  obtained using hydrodynamical methods. On the other hand, applying the approximate formula (\ref{lted}) for the energy density in equation (\ref{man}), we may estimate the eccentricity $\epsilon_{x}$ for regions $III$ and $III$$'$ of subsection \ref{IS}. We find

\label{Ecc1}
\begin{align}
\epsilon_{x}(\tau)\Big|_{b \ll \tau}&  \approx  \frac{\iint_{D_{III}\cup D_{III'}} dx^1dx^2 ((x^1)^2-(x^2)^2 )}{\iint_{D_{III}\cup D_{III'}} dx^1dx^2 ((x^1)^2+(x^2)^2)} \label{E2}
\end{align}
where $D_{III} \cup D_{III'}$ is the two dimensional surface defined in figure \ref{re} as the intersection of the circles of radius $\tau$ (almond shape). Thus the integration area is $\tau$ dependent. In principle, we should add the contributions to $\epsilon_x$ from the other regions. Unfortunately, performing the corresponding integrations (using $T_{00}$ from (\ref{SE})) analytically is hard and so is left for a future project. We argue that in a realistic situation of a heavy ion collision the important contribution to spatial eccentricity comes from the overlapping region (the almond) while whatever is left out is less important. Hence, equation (\ref{Ecc1}) may suffice as a first approximation in determining $\epsilon_x$. Assuming this and performing the integrations of (\ref{Ecc1}) and employing formula (\ref{lted}) we find that
\begin{align}
\label{EcIII}
\epsilon_{x}(\tau)\Big|_{b\ll \tau}  \approx \frac{2}{3 x^2} \frac{\sqrt{x^2-1}(1+2 x^2)-3 x^2 \sec^{-1}(x)}{-3\sqrt{x^2-1}+(2+x^2)\sec^{-1}(x)}, \hspace{0.3in}x=\frac{\tau}{b}.
\end{align}

\begin{figure}
\centering
\includegraphics*[scale=0.7,angle=0,clip=true]{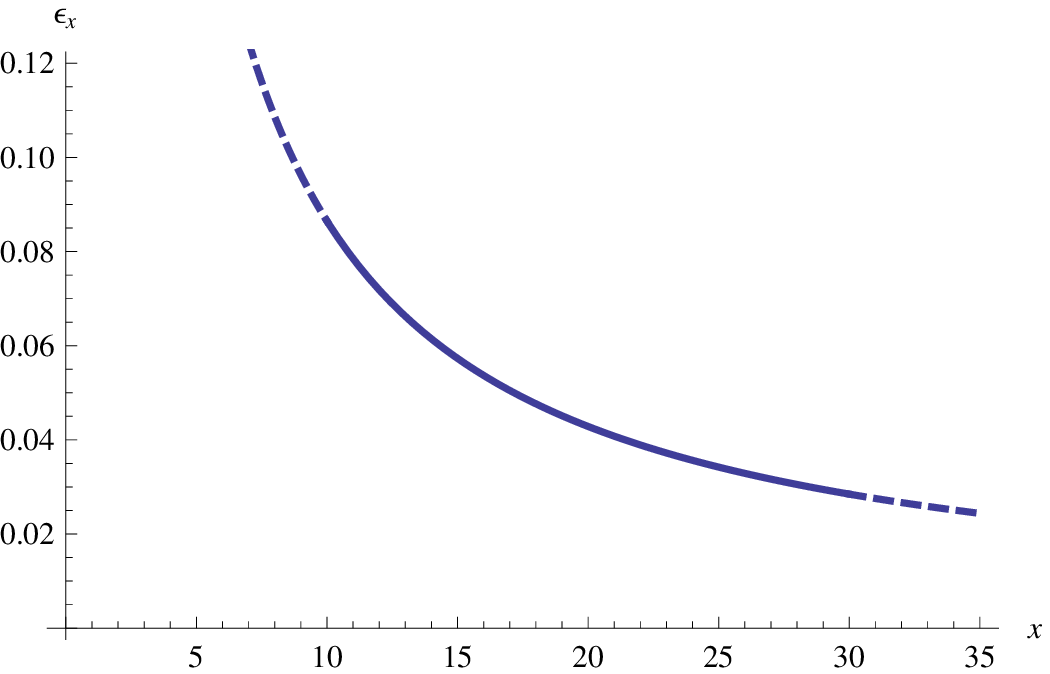} \hspace{1cm}
\includegraphics*[scale=0.7,angle=0,clip=true]{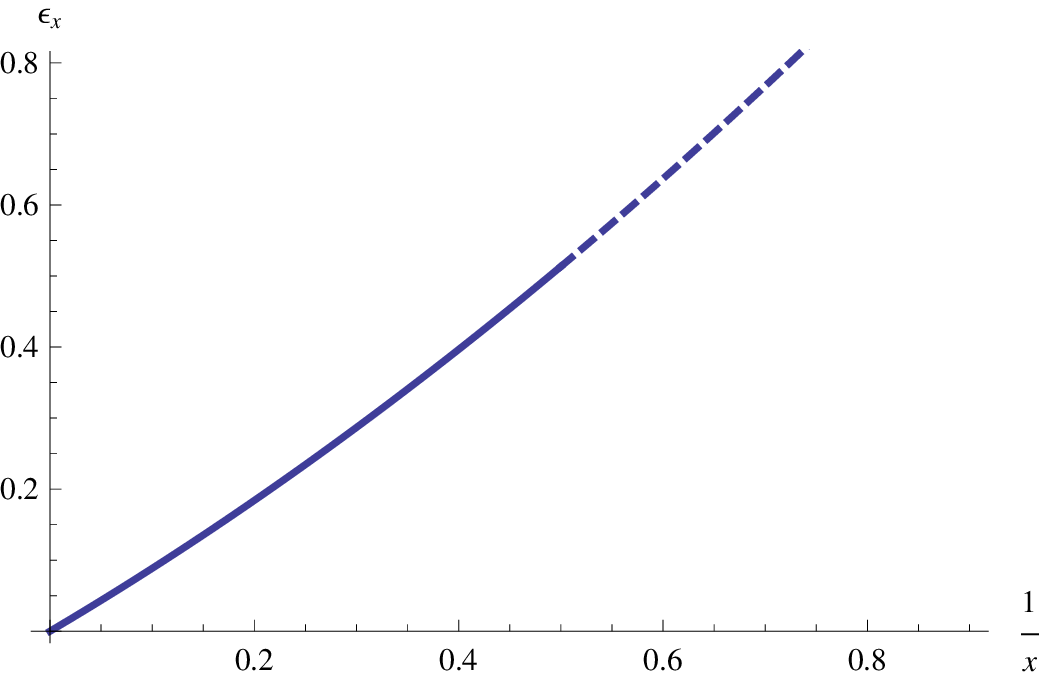}
\caption{The left plot depicts the eccentricity $\epsilon_x$ given by equation (7.23) as a function of $x=\frac{\tau}{b}$ for $x\gg 1$ (we assumed $\tau>10 \hspace{0.01in}b$). It approaches asymptotically to zero at large $x$. The right plot depicts $\epsilon_x$ (also from equation (7.23)) as a function of $\frac{1}{x}=\frac{b}{\tau}$ for (again) $x\gg 1$ (we assumed $\frac{1}{x}<0.5$). It provides the same information as the plot on the left and it implies that for fixed $\tau$ and as $b \rightarrow 0$, we have that $\epsilon_x \rightarrow  0$. In both figures, the dashed lines represent the asymptotic behavior of $\epsilon_x$ at regions where our approximations are no longer true as either $\tau$ is not large enough (compared to $b$ and hence (7.23) is not applicable) or because inequality (7.6) is violated (large $\tau$).}
\label{ecfig}
\end{figure}

Equation (\ref{EcIII}) shows that $\epsilon_{x}$ is a function of the ratio $x=\frac{\tau}{b}$ while the result (\ref{EcIII}) applies for $x\gg 1$. In figure \ref{ecfig} we plot $\epsilon_{x}$ as expressed in (\ref{EcIII}) as a function of $x$ and also as a function of $\frac{1}{x}$ for $x\geq 1$. For smaller values of $x$ equation (\ref{EcIII}) is not applicable. Nonetheless, this constraint can be easily removed and our result may be extended to smaller values of $x$ (that is $\tau$) by using the full result for $T_{\mu \nu}$ given by equation (\ref{SE}). This is left for future research (due to the complexity of (\ref{SE})). From the figure we see that as $\tau$ increases significantly compared to $b$ (left figure) or equivalently as $b \rightarrow 0$ for fixed $\tau$ (right figure), we have that $\epsilon_x \rightarrow 0$. The physics behind this is that when $\tau$ dominates over $b$, the eccentricity $\epsilon_x$ is insensitive to $b$ which expresses the asymmetry of the system and hence $\epsilon_x$ tends to zero. On the other hand as $x$ moves towards even larger values ($\tau$ increases), equation (\ref{E2}) is no longer valid because linear gravity is not applicable any more (that is inequality (\ref{mt}) is violated). In order to overcome this constraint, one needs to include higher order terms in $\mu$ in our perturbative solution. Nevertheless, figure \ref{ecfig} describes qualitatively eccentricity for these (intermediate) values of $x=\frac{\tau}{b}$  and it shares the same asymptotic behavior with results already obtained in the literature using pQCD methods \cite{Lappi:2006xc,Jas:2007rw} and hydrodynamic simulations \cite{Kolb:2000sd}, \cite{Kolb:2003dz,Kolb:2002cq}.


\subsection{Conclusions}\label{con}


Using the AdS/CFT correspondence we predict the evolution of the {\bf SE} tensor of the produced medium in heavy ion collisions taking into account a nontrivial transverse coordinate dependence. Our predictions apply for sufficiently early (see subsection \ref{va}) proper times. Our conclusions are summarized as follows.

\vspace{0.3in}

1. The {\bf SE} tensor $T_{\mu \nu}$ (to $O(\mu^2)$) is given by equation (\ref{SE}) and evolves nontrivially constrained by causality in an intuitive way. In particular, the behavior of $T_{\mu \nu}$ at any point on the transverse plane, is determined whether the propagation from the center of each individual nucleus has enough proper time to reach the point under consideration or not. Figure \ref{re} is a snapshot taken at given proper time $\tau$ and depicts the six kinematical regions where $T_{\mu \nu}$ evolves differently while it has the general form

\begin{align}\label{TMN}
T_{\mu \nu}^{(2)}&=\mu^2\theta(x^+)\theta(x^-)\Big\{ \theta(r_2-r_1)\theta(r_1-\tau)A_{\mu \nu}^I(x^{\kappa},b) + \theta(r_2-\tau)\theta(\tau-r_1)A_{\mu \nu}^{II}(x^{\kappa},b)  \notag\\&
 +\theta(r_2-r_1)\theta(\tau-r_2)A_{\mu \nu}^{III}(x^{\kappa},b) + (b \leftrightarrow -b) \Big \}.
\end{align}
The indices $I$, $II$, $III$ on $A_{\mu \nu}$ correspond to the regions  $I$, $II$, $III$ of figure \ref{re} respectively \footnote{The  $(b \leftrightarrow -b)$ terms cover the rest three remaining $I$$'$, $II$$'$ and $III$$'$ since under this interchange $r_1 \leftrightarrow r_2$ .  }.

\vspace{0.3in}

2. Our early (proper) time analysis (see subsection (\ref{va})) is not adequate to prove or disprove wether a thermal medium is produced allowing for its hydrodynamical description. In principle, one could get a hint  for the formation of a horizon and hence for thermalization by plotting the (components of the) metric $g_{MN}$ up to the order $\mu^2$, which is what has been computed in this work \footnote{Actually in this paper we have only computed the $z^4$ coefficient of $g_{MN}$  to $O(\mu^2)$. Computing the rest powers in $z$ (to $O(\mu^2)$) may be achieved similarly.}, and investigate the possibility for the metric to change its sign on some (generally) hypersurface \footnote{This hypersurface, if exists, may be in principle found by an analytic computation. Plotting the metric would be a convenient way to locate the hypersurface but it is not a necessary step.}. If this is the case, then the set of points on this hypersurface would correspond to some temperature of the medium which would generally depend on $x^{\mu},\mu$ and $b$. Such a procedure could give an estimation for the thermalization time; this is non-trivial because of the complication of the corresponding formulas and it is therefore left for future investigation. However, even if a change in the sign of the metric appears, this wouldn't suffice to prove thermalization but it would merely be a good indication that it occurs. This is because of the incomplete analysis, that is the linear approximation, which  we have used in the present work. In order to have a complete proof one should solve the problem to all orders in $\mu$, that is the full non-linear problem.


\vspace{0.3in}

3. The energy density simplifies at $r_i \gg \tau$, i=1,2 and $r_i \ll \tau$ where $r_i$ is the distance of the arbitrary point $\vec{r}$ on the transverse plane, from the center of the nucleus i where i=1,2. It behaves in a way proportional to the product of the {\bf SE} tensor of the two nuclei. In particular for these two regions ($r_i \gg \tau$ and $r_i \ll \tau$) we get that the energy density evolves as in (\ref{eted}) and (\ref{lted}) respectively. More specifically, the energy density for $r_i \ll \tau$ has a logarithmic dependence with $\tau$. Our result has the same $\log^2( k\tau)$ dependence as  \cite{Lappi:2003bi} (see also \cite{Fukushima:2007ja,Fries:2006pv}) which has been derived using pQCD. However, our result (at central rapidities) has in addition an overall $\tau^2$ dependence. Thus one may conclude that (in the aforementioned kinematical window and in the limit $b \rightarrow 0$) the energy density, except from an overall $\tau^2$ dependence, is symmetric under $\tau \leftrightarrow r$. In fact, this conclusion is applicable to all the regions where our analytic formula for $T_{\mu \nu}$, equation (\ref{TMN}), is applicable. A similar behavior has been observed in \cite{Gubser:2010ze}.
\vspace{0.3in}

4. The formula we found for $T_{\mu \nu}$ (see (\ref{SE})) and which is valid at early proper times allows us to compute the momentum anisotropy and spatial eccentricity as a function of the ratio $\frac{\tau}{b}$. In this paper, we perform the calculations for both quantities for $\tau \gg b$ \footnote{But still sufficiently at small $\tau$.}. We find that our results (see (\ref{MA}), (\ref{EcIII}) and figure \ref{ecfig}) agree qualitatively with results in the literature obtained in the context of pQCD \cite{Lappi:2006xc,Jas:2007rw} and hydrodynamic simulations \cite{Kolb:2000sd}, \cite{Kolb:2003dz,Kolb:2002cq}.

\vspace{0.3in}

5. We have seen that the {\bf SE} tensor of the produced medium depends strongly on the product of the  {\bf SE} tensor of each nucleus before the collision. What is more interesting is that when the proper time is large compared to the observational point, the argument $r_i$ of the profile is replaced by $\tau$. We conjecture that this conclusion is independent of the choice of the transverse profiles under the logical hypothesis that the nuclear density is greater in the center of the nucleus than the edges. Assuming this in a more realistic situation, one may take the Woods-Saxon energy density of a nucleus at rest given by

 \begin{align}\label{WS1}
T_{00}=\epsilon_{WS}(r,x^3) \approx M \frac{1}{1+e^{\frac{|\vec{x}|-R}{\alpha}}}, \hspace {0.3in} \vec{x}=(x^1,x^2,x^3)
\end{align}
where $\alpha$ and R are positive constants and M has mass dimension four  (compare with $\mu$ that we used in this paper) while the remaining components of $T_{\mu \nu}$ are zero. The next step is to boost this tensor (see also \cite{Gubser:2008pc}, \cite{Aichelburg:1970dh} and \cite{Hotanaka:1993}) into the infinite momentum frame keeping the ratio $\mu=\frac{M}{\sqrt{1-\beta^2}}$, where $\frac{1}{\sqrt{1-\beta^2}}$ is the boost factor, fixed. Assuming the boost is along $x^3$, the result of such a boost gives

\begin{align}\label{WS2}
T_{--}=\mu \delta(x^-) \int^{\infty}_{-\infty}dx^3 \frac{1}{1+e^{\frac{\sqrt{r^2+(x^3)^2}-R}{\alpha}}}, \hspace{0.3in}r^2=(x^1)^2+(x^2)^2.
\end{align}

Consequently, according to our conjecture, we believe that at early times the energy density should depend strongly on the factor

\begin{align}\label{WS3}
\int^{\infty}_{-\infty}dx^3 \epsilon_{WS}( \eta _>,x^3) \int^{\infty}_{-\infty}dx^3 \epsilon_{WS}( \xi _>,x^3), \hspace{0.3in}   \xi _>=max\{r_1,\tau\},\hspace{0.05in}\eta _>=max\{r_2,\tau \} 
\end{align}
where $r_{i}$ is the distance of the observational point from the center of the nucleus i (i=1,2). Hence at the center of the collision where the proper time $\tau$ satisfies $\tau>r_{1,2}$, we conjecture that the energy density (and generally $T_{\mu \nu}$) behaves like

\begin{align}\label{WS}
\epsilon(\tau)  \sim  \left(\mu \tau \int^{\infty}_{-\infty}dx^3 \frac{1}{1+e^{\frac{\sqrt{\tau^2+(x^3)^2}-R}{\alpha}}} \right)^2.
\end{align}

\vspace{0.3in}

6. Any terms in equation (\ref{SE}) that give $T_{\mu \nu}$ not proportional to ${\cal J}$, ${\cal M}^{(0)}$ and ${\cal M}^{(2)}$ are due to $J_{MN}$. As is evident from these terms, the presence of the bulk sources $J_{MN}$ and the back-reactions affect $g_{\mu \nu}$ and hence $T_{\mu \nu}$ not only on the forward light-cone but also inside. This implies that we cannot in principle solve Einstein's equations in vacuum arguing that we are away from the sources unless we know the boundary conditions that these sources enforce on the metric. This is in analogy to classical electrodynamics: solving Laplace equation for the scalar potential away from a point charge sitting at the origin without specifying the boundary conditions, one may obtain the trivial (zero) solution which obviously is not the correct one.

\vspace{0.3in}

7. The presence of the impact parameter $b$ is a necessary requirement and not an additional complication introduced in the problem. This is obvious from the solution of  the {\bf SE}, equation (\ref{SE}), tensor which diverges in the limit $b \rightarrow 0$. More importantly in the same limit, according to (\ref{Jcon}), the conservation of $J_{MN}$ is violated.
This suggests that a head-on
 collision may not be investigated using classical gravity. Instead, one has to apply a quantum theory of gravity in the same way one can not predict the electron-positron annihilation (in head-on
 collisions) using Maxwell's equations. One has to turn to Quantum Electrodynamics in order to describe the process.

\vspace{0.3in}

8. We still face the problem of negative energy densities appearing in \cite{Kajantie:2007bn} - \cite{Grumiller:2008va}. This is implied by the fact that while conservation (see (\ref{ec})) applies, the initial shock waves continue to propagate in the forward light cone unaltered. Since these carry the total energy of the collision, we conclude that the net energy production should add to zero. This should not be a surprise but rather a feature of our approximation; it is just a consequence of the eikonal approximation which we consider in this work. Since the eikonal approximation is always valid at early times (see \cite{Kovchegov:1997ke} for some eikonal results), negative energy is always present.

\vspace{0.3in}

9. For future projects we propose that one could compute the spatial eccentricity and momentum anisotropy at early times extending our preliminary results (see conclusion 3) by making use of the (full) analytic formula (\ref{SE}) that computes $T_{\mu \nu}$ (for sufficiently early times). This would extend the result of the left plot of figure \ref{ecfig} to smaller values of $x=\frac{\tau}{b}$ (that is to the left). In addition, a similar analysis could be applied to the shock waves that correspond to (\ref{g8}) (see \cite{Gubser:2008pc} and \cite{Gubser:2009sx}). These shock waves have the  advantage that they are created by a point-like {\bf SE} bulk tensor and hence the back-reaction should be specified by the geodesics. Unfortunately, their form is very complicated and the analysis would be much more involved. 

\vspace{0.3in}

10. Finally, our technique has already been applied to shock wave collisions in ordinary four dimensional  gravity \cite{Taliotis:2010az} , taking into account back-reactions. The problem of axisymmetric  collisions in four dimensional gravity has been analyzed extensively while the solutions were obtained in vacuum \cite{DEath:1992hb,DEath:1992hd,DEath:1992qu}. In \cite{Taliotis:2010az} one may see the way solution of \cite{DEath:1992hb,DEath:1992hd,DEath:1992qu} gets modified in the presence of matter and non-zero impact parameter.

\acknowledgments

The author would like to thank Huichao Song, Ulrich Heinz, Zhi Qiu, Kyle Wendt, Edmond Iancu and Yuri Kovchegov for stimulating discussions and Ben Dundee and Eric Braaten for reading the manuscript. He also thanks the organizers of TASI 2010 at Boulder, Colorado  for their excellent hospitality during the writing of a part of this work. In addition, he owns gratefulness to George Bruhn, Oscar Chacaltana, Alberto Faraggi, Idse Heemskerk, Philip Szepietowski and Andreas Stergiou  for the endless and inspiring discussions about physics at the $Goose$ at Boulder during their participation at TASI. Special thanks to Andreas Stergiou for making valuable commends and suggestions about this project.  
Finally, he would like to thank Yuri Kovchegov in particular  for his continuous encouragement. This work is sponsored in part by the U.S. Department of Energy under Grant No. DE-FG02-05ER41377 and in part by the Institution of Governmental Scholarships of Cyprus (IKY).


\appendix
\renewcommand{\theequation}{A\arabic{equation}}
\setcounter{equation}{0}
\section{Solving equations (5.8) }
\label{A}
In this appendix we perform the $x^{\pm}$ part of the integrations resulting when the Green's function acts on the right-hand side of (\ref{deq}). Defining
\begin{align}\label{ro}
\rho=|\vec{r}\hspace{0.02in}'-\vec{r}|
\end{align}
and proceeding as in (\ref{Got}) we find out that we have to deal with five different cases:

\vspace{0.3in}
\hspace{0.25in}\underline{Case I: $z^n\delta(x^+)\delta(x^-)$ terms}
\vspace{0.3in}
\begin{subequations}\label{c1}
\begin{align}
 &G \otimes z'^6\delta(x'^+)\delta(x'^-)f(\vec{r}\hspace{0.02in}')=3 z^4\theta(x^+)\theta(x^-)\frac{1}{2 \pi}\int d^2\vec{r} \theta(\tau-\rho) f(\vec{r}\hspace{0.02in}'),
\label{c1a}\\&
G \otimes z'^8\delta(x'^+)\delta(x'^-)f(\vec{r}\hspace{0.02in}')=12          z^4\theta(x^+)\theta(x^-)\frac{1}{2 \pi}\int d^2\vec{r} \theta(\tau-\rho)(\tau^2-\rho^2) f(\vec{r}\hspace{0.02in}').\label{c1b}
\end{align}
\end{subequations}
\vspace{0.3in}
\hspace{0.25in}\underline{Case II: $z^8\delta'(x^+)\delta(x^-)$ terms}
\vspace{0.3in}
\begin{align}\label{c2}
G \otimes z'^8\delta'(x'^+)\delta(x'^-)f(\vec{r}\hspace{0.02in}')=24x^-z^4\theta(x^+)\theta(x^-)\frac{1}{2 \pi}\int d^2\vec{r} \theta(\tau-\rho) f(\vec{r}\hspace{0.02in}'),
\end{align}
where we have integrated by parts the $\delta(x'^+)$ and have dropped two vanishing terms.

\vspace{0.3in}
\hspace{0.25in}\underline{Case III: $z^n (x^-)^m\delta(x^+)\theta(x^-)$ terms}
\vspace{0.3in}
\begin{subequations}\label{c3}
\begin{align}
&G \otimes z'^8\delta(x'^+)\theta(x'^-)f(\vec{r}\hspace{0.02in}')=\frac{3}{\pi}\frac{x^-}{\tau^2}z^4\theta(x^+)\theta(x^-)\int d^2\vec{r} \theta(\tau-\rho)(\tau^2-\rho^2)^2 f(\vec{r}\hspace{0.02in}')\label{c3a}\\&
G \otimes z'^6 x'^-\theta(x'^-)\delta(x^+)f(\vec{r}\hspace{0.02in}')=\frac{3}{2}\frac{(x^-)^2}{\tau^4}z^4\theta(x^+)\theta(x^-)\frac{1}{2 \pi}\int d^2\vec{r} \theta(\tau-\rho)(\tau^2-\rho^2)^2 f(\vec{r}\hspace{0.02in}')\label{c3b}
\end{align}
\end{subequations}
where in both cases we have used the identity
\begin{align}\label{thetaid}
&\hspace{0.23in}\theta(x^+)\int_{-\infty}^{\infty}dx'^- \theta(x^- - x'^-)\theta(\sqrt{2 x^+(x^- - x'^-)}-\rho)\theta(x'^-)f(x'^-)\notag\\&
=\theta(x^+)\theta(x^-)\int_{0}^{x^-}dx'^-\theta(\sqrt{2 x^+(x^- - x'^-)}-\rho)f(x'^-)\notag\\&
=\theta(x^+)\theta(x^-)\theta(\tau-\rho)\int_{0}^{\frac{\tau^2-\rho^2}{2 x^+}}dx'^-f(x'^-)
\end{align}
\vspace{0.3in}
\hspace{0.25in}\underline{Case IV: $z^8 \delta''(x^+)\delta(x^-)$ terms}
\begin{align}\label{c4}
G \otimes z'^8\delta''(x'^+)\delta(x'^-)f(\vec{r}\hspace{0.02in}')&=\frac{6}{\pi}z^4\theta(x^-)    \left(\partial^2_{x^+}\right) \Bigg \{  \int_{-\infty}^{\infty}dx'^+ \int d^2\vec{r} \hspace{0.02in}'   \delta(x'^+) \theta\left(x^+ - x'^+\right)\notag\\&
\hspace{0.22in} \times \theta(\sqrt{2(x^+-x'^+)x^-}-\rho) \left(2 (x^+-x'^+)x^--\rho^2 \right)\Bigg \} f(\vec{r}\hspace{0.02in}')\notag\\&
=\frac{6}{\pi}z^4  \theta(x^-)(2x^-)    \left(\partial_{x^+}\right) \Big\{\theta(x^+)    \int d^2\vec{r}   \theta(\sqrt{2x^+x^-}-\rho^2)\Big \}f(\vec{r}\hspace{0.02in}')\notag\\&
=24(x^-)^2z^4  \theta(x^+)\theta(x^-)\frac{1}{2 \pi \tau} \int d^2\vec{r} \hspace{0.02in}' \delta(\tau-\rho) f(\vec{r}\hspace{0.02in}').
\end{align}
A few explanations are in order: In the first equality we integrated by parts twice the $\delta$-function and exchanged $\partial_{x'^+} \leftrightarrow -\partial_{x^+}$ (these act on the curly bracket), in the second equality we performed the $x'^+$ integration and the first derivative with respect to $x^+$ ignoring two vanishing terms and in the third equality we performed the second differentiation ignoring one more vanishing term.

\vspace{0.3in}
\hspace{0.25in}\underline{Case V: $z^8 x^-\theta(x^-)\delta(x^+) \left( t_{2,x^1}  \nabla^2_{\perp} t_{1,x^1} + t_{2,x^2}  \nabla^2_{\perp} t_{1,x^2} \right)$}
\vspace{0.1in}

As mentioned earlier in subsection \ref{TP} the expressions $t_1$ and $t_2$ carry only the transverse dependence. We have

\begin{align}\label{c5a}
G & \otimes  z'^8  x'^-\theta(x'^-)\delta(x'^+)  \left( t_{2,x'^1}  \nabla^2_{\perp} t_{1,x'^1} + t_{2,x'^2}  \nabla^2_{\perp} t_{1,x'^2} \right)\notag\\&
=\frac{6}{\pi}z^4  \theta(x^+)\int_{-\infty}^{\infty}dx'^- \int d^2\vec{r} \hspace{0.02in}'x'^-  \theta(x'^-)  \theta(x^- - x'^-)   \theta(\sqrt{2x^+(x^--x'^-)}-\rho)\notag\\&
\hspace{1.5in} \times \left(2 x^+(x^--x'^-)-\rho^2 \right)
\left( t_{2,x'^1}  \nabla^2_{\perp} t_{1,x'^1} + t_{2,x'^2}  \nabla^2_{\perp} t_{1,x'^2} \right)\notag\\&
=\frac{1}{\pi} z^4  \frac{(x^-)^2}{\tau^4}\theta(x^+)\theta(x^-)\int d^2\vec{r} \hspace{0.02in}'\theta\left(\tau-\sqrt{(x^1-x'^1)^2+(x^2-x'^2)^2}\right) \notag\\&
\hspace{1.4in}\times \left(\tau^2- \left((x^1-x'^1)^2+(x^2-x'^2)^2\right)\right)^3\left( t_{2,x'^1}  \nabla^2_{\perp} t_{1,x'^1} + t_{2,x'^2}  \nabla^2_{\perp} t_{1,x'^2} \right),
\end{align}
where in the first equality we performed the $x^+$ integration while in the second equality we performed the $x^-$ integration with the help of (\ref{thetaid}) and also we substituted (\ref{ro}). In order to perform the transverse integrations we integrate (once) by parts the term $\nabla^2_{\perp} t_{1,x'^1}=\partial_{x'^1}(\nabla^2_{\perp} t_1)  $ with respect to $x'^1$. Ignoring the surface term because of the $\theta$-function, we obtain three terms. One term contains a $\delta$-function term which vanishes (it has the form $x\delta(x)$). The second term is proportional to $t_{2,x'^1x'^1}  \nabla^2_{\perp} t_1$ which together with the integration by parts of the term $\nabla^2_{\perp} t_{1,x'^2}$ with respect to $x'^2$ gives the contribution $\sim \nabla^2_{\perp} t_2 \nabla^2_{\perp} t_1$. But this term, for nonzero impact parameter, is according to (\ref{nab^2}) zero. Hence, we are left with the term where $\partial_{x'^1}$ acts on $\left(\tau^2- \left((x^1-x'^1)^2+(x^2-x'^2)^2\right)\right)^3$ and with a similar term from the action of $\partial_{x'^2}$. Thus, the integrations by parts simplifies to an integral proportional to $\nabla_{\bot}^2 t_1$ which according to (\ref{nab1}) results to $\sim \delta(\vec{r}\hspace{0.02in}-\vec{b}_1)$ and so the integrations become trivial. The final result is then
\begin{align}\label{c5}
G & \otimes  z'^8  x'^-\theta(x'^-)\delta(x'^+)  \left( t_{2,x'^1}  \nabla^2_{\perp} t_{1,x'^1} + t_{2,x'^2}  \nabla^2_{\perp} t_{1,x'^2} \right)\notag\\&
=-12 \mu^2 z^4  \theta(x^+)\theta(x^-)\frac{(x^-)^2}{\tau^4 }\frac{(\tau^2-r_1^2)^2}{|\vec{b}_1-\vec{b}_2|^2}\left( (b_{11}-b_{21})(x_1-b_{11})+(b_{12}-b_{22})(x_2-b_{12}) \right)
\end{align}
where the $b_{ij}$'s are defined in (\ref{b}).
\renewcommand{\theequation}{C\arabic{equation}}
\setcounter{equation}{0}
\section{Evaluating the integral (6.16) }
\label{C}

In this appendix we calculate the two integrals ${\cal M}^{(n)}$ of (\ref{M}) for $n=0$ and $n=2$. We begin with the ${\cal M}^{(0)}$ integral first

\begin{align} \label{Ib}
{\cal M}^{(0)}=\frac{1}{2\pi}\int d^2r' (\theta(\tau-|\vec{r}-\vec{r'}|) \log(kr'_1)\log(k r'_2).
\end{align}
The quantities $r_{1,2}$ are given by (\ref{r12}). The area of integration is a circle of radius $\tau$ centered at $r$. The trick here is to expand the logarithms in their Fourier space: $\log(kr)=-\int \frac{d^2q}{2\pi}\frac{ e^{i \vec{q} \hspace{0.02in} \vec{r}}}{q^2}$ with $k$ serving as an ultraviolet cutoff.
Shifting the variable $\vec{r'}\rightarrow\vec{r'}+\vec{r}$, expanding both logarithms and performing the angular integration one obtains \footnote{Recall from (\ref{r12}) that $\vec{r_i}=\vec{r} \mp \vec{b}$, $ i=1,2$.}

\begin{align}\label{Ic}
{\cal M}^{(0)}= \int_{0}^{\tau}dr' r'\left \{ \int\frac{d^2q d^2l}{(2\pi)^2} \frac{e^{i \vec{q} \hspace{0.02in} \vec{r_1}+i \vec{l} \hspace{0.02in} \vec{r_2} }} {q^2 l^2} J_0 (r'|\vec{q}+\vec{l}|) \right\} \equiv  \int_{0}^{\tau}dr' r'{\cal J}(r').
\end{align}

Next we perform the $l$ and $q$ integrations. The integrals that define ${\cal J}$ have been calculated in \cite{Kovchegov:1997ke}. In order to perform these integrations one has to expand $ J_0 (r'|\vec{q}+\vec{l}|)$ in an infinite sum of products of the form $ J_n (r'|\vec{q}|) J_n (r'|\vec{l}|)$ with $n$ an integer and do the angular integrals (of q and l) first. This factors out the radial integrations over $l$ and $q$ into two independent integrals. Then one has to perform these integrations and finally sum over $n$. The final result reads

\begin{subequations}\label{Ja}
\begin{align}
{\cal J}(r')& \equiv \ln(\xi_> k)\ln(\eta_> k)+\frac{1}{4}\left[ Li_2\left( e^{i\alpha}\frac{\xi_<\eta_<}{\xi_>\eta_>} \right) + Li_2\left( e^{-i\alpha}\frac{\xi_<\eta_<}{\xi_>\eta_>} \right)\right], \label{J}\\
\xi_{>(<)}&=max(min)(r_1,r') \hspace{0.2in} \eta_{>(<)}=max(min)(r_2,r'), \label{ke}\\
& \hspace{1.2in} (\vec{r_1}).(\vec{r_2})=\cos(\alpha) r_1 r_2. \label{a} 
\end{align}
\end{subequations}
Here $\alpha$ is the angle between $\vec{r}_1$ and $\vec{r}_2$ and $Li_2$ is the dilogarithm function. We also note that ${\cal J}$ is real as it should.

We still have to integrate (\ref{J}) over $r'$ in order to obtain ${\cal M}^{(0)}$. This is not trivial since according to the definition of $\xi_{>(<)}$ and $\eta_{>(<)}$, we have to break the $r'$ integration into cases. In practice, we have to introduce $\theta$ functions. We organize the several cases by first introducing table \ref{ta1} \footnote{${\cal J}_3$ for instance means ${\cal J}(\xi_>\eta_>=r'r_2,\xi_<\eta_<=r'r_1)$ with ${\cal J}$ given from (\ref{J}). }. With the help of the table and taking into account all the six possible ways one can order $r_1$, $r_2$ and $\tau$ it is found that

\begin{table}
\caption{We defined $\vec{r}_{1,2}=(x^1 \mp b,x^2)$}
\centering
\begin{tabular}{c|cccc|r}
\hline\hline
cases & $\xi_>$ & $\eta_>$ & $\xi_>\eta_>$ & $\xi_<\eta_<$ & ${\cal J}_i$ \\
\hline\hline
1 & $r_1$ & $r_2$ & $r_1r_2$ & $r'^2$ & ${\cal J}_1$ \\
2 & $r_1$ & $r'$ & $r'r_1$ & $r'r_2$ & ${\cal J}_2$\\
3 & $r'$ & $r_2$ & $r'r_2$ & $r'r_1$ & ${\cal J}_3$ \\
4 & $r'$ & $r'$ & $r'^2$ & $r_1r_2$ & $ {\cal J}_4$ \\
\hline\hline
\end{tabular}
\label{ta1}
\end{table}

\begin{align}\label{intJ}
{\cal M}^{(0)}= \int_{0}^{\tau}dr'r'{\cal J}&= 
\theta(r_2-r_1)\theta(\tau-r_2) \left[\int_{0}^{r_1}dr' r'{\cal J}_1 + \int_{r_1}^{r_2}dr' r'{\cal J}_3 + \int_{r_2}^{\tau}dr' r'{\cal J}_4 \right] \notag\\& 
+\theta(r_1-r_2)\theta(\tau-r_1) \left[ \int_{0}^{r_2}dr' r'{\cal J}_1 + \int_{r_2}^{r_1}dr' r'{\cal J}_2 + \int_{r_1}^{\tau}dr' r'{\cal J}_4 \right] \notag\\& 
+ \theta(\tau-r_1)\theta(r_2-\tau) \left[ \int_{0}^{r_1}dr' r'{\cal J}_1 + \int_{r_1}^{\tau}dr' r'{\cal J}_3 \right]\notag\\&
+ \theta(\tau-r_2)\theta(r_1-\tau) \left[ \int_{0}^{r_2}dr' r'{\cal J}_1 + \int_{r_2}^{\tau}dr' r'{\cal J}_2 \right]\notag\\&
+\left( \theta(r_2-r_1)\theta(r_1-\tau) +\theta(r_1-r_2)\theta(r_2-\tau) \right ) \int_{0}^{\tau}dr' r'{\cal J}_1 .
\end{align}
The integrals involving ${\cal J}_2$, ${\cal J}_2$ and ${\cal J}_3$ are easy to calculate. The $ {\cal J}_4$ integral is harder. It involves integrations of the form $\int d'r r' Li_2\left( e^{\pm i\alpha}\frac{r_1 r_2}{r'^2} \right) $. In order to do such integrals we use the integral representation of $Li_2$
\begin{align}\label{L2a}
Li_2(z)=z\int_{0}^{\infty}\frac{w}{e^w-z}dw.
\end{align}
Then we perform the integrals by first integrating over $r'$ and then over the parameter $w$ as follows

\begin{align}\label{L2b}
\hspace{0.4in} & \int d'r r' Li_2\left( e^{ i\alpha}\frac{r_1 r_2}{r'^2} \right) = \notag\\&
\int_{0}^{\infty}\left[ \int dr' r' e^{ i\alpha}\frac{r_1 r_2}{r'^2} \frac{w}{e^w- e^{ i\alpha}\frac{r_1 r_2}{r'^2}} \right]dw= \frac{r_1 r_2}
{2}e^{ \alpha} \int _{0}^{\infty} w e^{-w} \log(r'^2 e^w-r_1r_2 e^{ i \alpha} ) dw=\notag\\&
\frac{1}{2}\left[ \left(e^{i\alpha}r_1 r_2 -r'^2 \right) \log \left( 1-\frac{e^{i\alpha}r_1 r_2}{r'^2} \right) + e^{i\alpha}r_1 r_2 \log \left( \frac{r'^2}{e^{i\alpha} r_1 r_2} \right) + r'^2 Li_2\left( \frac{e^{i\alpha}r_1 r_2}{r'^2} \right) \right].
\end{align}
Verifying that the derivative of the right-hand side of (\ref{L2b}) is equal to the integrand is straightforward. Now using this result, we may obtain the results of all integrations whose values in their indefinite form (up to a constant) are 

\begin{subequations}\label{Ii}
\begin{align}
&{\cal M}_1^{(0)}(r')\equiv \int dr' r'{\cal J}_1 =\frac{ r'^2}{2} \log(k r_1) \log(k r_2) + \frac{1}{8}\Bigg[ - e^{i\alpha} r_1 r_2 \log \left(1- e^{-i\alpha} \frac{r'^2}{r_1 r_2} \right)\notag\\&
\hspace{1.3in} +r'^2 \left(-1+ \log \left( 1-e^{-i\alpha} \frac{r'^2}{r_1 r_2} \right) + Li_1\left( e^{-i\alpha}\frac{r'^2}{r_1 r_2} \right) \right) + c.c. \Bigg] ,  \label{I1}\\
&{\cal M}_2^{(0)}(r') \equiv \int dr'r'{\cal J}_2 = \frac{r'^2}{2} \log(k r_1) \left( -\frac{1}{2} + \log(k r') \right) 
+ \frac{r'^2}{8} \left[ Li_2\left( e^{i\alpha}\frac{r_2}{r_1 } \right) + c.c. \right] ,   \label{I2}\\
&{\cal M}_3^{(0)}(r') \equiv \int dr' r'{\cal J}_3 = \frac{r'^2}{2} \log(k r_2) \left( -\frac{1}{2} + \log(k r') \right) + \frac{r'^2}{8} \left[ Li_2\left( e^{i\alpha}\frac{r_1}{r_2 } \right) + c.c. \right] ,  \label{I3}\\
&{\cal M}_4^{(0)}(r') \equiv \int dr' r'{\cal J}_4 = \frac{r'^2}{2} \left( \frac{1}{2} + \left(\log(k r')\right)^2 -\log(k r') \right)
+\frac{1}{8}\Bigg[ \left(e^{i\alpha}r_1 r_2 -r'^2 \right) \log \left( 1-\frac{e^{i\alpha}r_1 r_2}{r'^2} \right)\notag\\&
\hspace{1.3in} + e^{i\alpha}r_1 r_2 \log \left(e^{-i\alpha} \frac{ r'^2}{r_1 r_2} \right) + r'^2 Li_2\left( \frac{e^{i\alpha}r_1 r_2}{r'^2} \right) +c.c. \Bigg ] .  \label{I4}
\end{align}
\end{subequations}
With these integrals at hand and taking into account that ${\cal M}^{(2)}_1(0)=0$, we write a final expression for ${\cal M}^{(0)}$. Using (\ref{intJ}) we eventually have

\begin{align}\label{I}
{\cal M}^{(0)}&= \theta(r_2-r_1)\theta(\tau-r_2) \left[{\cal M}_1^{(0)}(r_1)+ {\cal M}_3^{(0)}(r_2)-{\cal M}_3^{(0)}(r_1) + {\cal M}_4^{(0)}(\tau)-{\cal M}_4^{(0)}(r_2) \right] \notag\\& 
+\theta(r_1-r_2)\theta(\tau-r_1) \left[ {\cal M}_1^{(0)}(r_2) + {\cal M}_2^{(0)}(r_1)-{\cal M}_2^{(0)}(r_2) + {\cal M}_4^{(0)}(\tau)-{\cal M}_4^{(0)}(r_1) \right] \notag\\& 
+ \theta(\tau-r_1)\theta(r_2-\tau) \left[ {\cal M}_1^{(0)}(r_1)+ {\cal M}_3^{(0)}(\tau)-{\cal M}_3^{(0)}(r_1) \right ]\notag\\&
+ \theta(\tau-r_2)\theta(r_1-\tau) \left[ {\cal M}_1^{(0)}(r_2)+ {\cal M}_2^{(0)}(\tau)-{\cal M}_2^{(0)}(r_2) \right]\notag\\&
+\left( \theta(r_2-r_1)\theta(r_1-\tau) +\theta(r_1-r_2)\theta(r_2-\tau) \right ) {\cal M}_1^{(0)}(\tau).
\end{align}
with ${\cal M}^{(2)}_i$'s given by (\ref{Ii}).


We now proceed in evaluating the integral (\ref{M}) for $n=2$. The ${\cal M}^{(2)}$ integral is calculated as before: one has to again break the integration into cases and so on, repeating the steps from equations (\ref{Ib})-(\ref{I}). The difference now is that the factor $r'$ in the ${\cal M}^{(0)}$ integral of (\ref{Ib}) (and so for the integrals of (\ref{Ii})) is replaced by $r'(\tau^2-r'^2)$. Therefore, working exactly as before, one obtains

\begin{align}\label{K}
{\cal M}^{(2)}&= \theta(r_2-r_1)\theta(\tau-r_2) \left[{\cal M}_1^{(2)}(r_1)+ {\cal M}_3^{(2)}(r_2)-{\cal M}_3^{(2)}(r_1) + {\cal M}_4^{(2)}(\tau)-{\cal M}_4^{(2)}(r_2) \right] \notag\\& 
+\theta(r_1-r_2)\theta(\tau-r_1) \left[ {\cal M}_1^{(2)}(r_2) + {\cal M}_2^{(2)}(r_1)-{\cal M}_2^{(2)}(r_2) + {\cal M}_4^{(2)}(\tau)-{\cal M}_4^{(2)}(r_1) \right] \notag\\& 
+ \theta(\tau-r_1)\theta(r_2-\tau) \left[ {\cal M}_1^{(2)}(r_1)+ {\cal M}_3^{(2)}(\tau)-{\cal M}_3^{(2)}(r_1) \right ]\notag\\&
+ \theta(\tau-r_2)\theta(r_1-\tau) \left[ {\cal M}_1^{(2)}(r_2)+ {\cal M}_2^{(2)}(\tau)-{\cal M}_2^{(2)}(r_2) \right]\notag\\&
+\left( \theta(r_2-r_1)\theta(r_1-\tau) +\theta(r_1-r_2)\theta(r_2-\tau) \right ) {\cal M}_1^{(2)}(\tau).
\end{align}
The ${\cal M}_i^{(2)}$'s are analogous to those of (\ref{Ii}) and are associated with an analogous table as table \ref{ta1}. Their explicit forms are given by

\begin{subequations}\label{Ki}
\begin{align}
&{\cal M}_1^{(2)}(r')= \frac{ r'^2}{4} (2\tau^2-r'^2) \log(k r_1) \log(k r_2) + \frac{1}{64} \Bigg [ r'^4 - 8r'^2\tau'^2 \notag\\&
\hspace{0.5in}+2\left( e^{2i\alpha}r_1^2 r_2^2 -r'^4 +4r'^2\tau'^2 \right) \log \left( 1-e^{-i\alpha}\frac{r'^2}{r_1 r_2}\right) +2 e^{i\alpha}r_1 r_2 \notag\\&
\hspace{0.5in} \times \left(r'^2-4\tau^2 \log \left( 1-e^{-i\alpha}\frac{r'^2}{r_1 r_2}\right) \right)
+4r'^2(2\tau^2-r'^2) Li_2\left( e^{-i\alpha}\frac{r'^2}{r_1 r_2} \right) + c.c. \Bigg] \label{K1}\\
&{\cal M}_2^{(2)}(r') = \frac{r'^2}{16} \log(k r_1) \left(r'^2-4\tau^2+4(2\tau^2-r'^2) \log(k r') \right) + \frac{1}{16}r'^2 (2\tau^2-r'^2) \left[ Li_2\left( e^{i\alpha}\frac{r_2}{r_1 } \right)+c.c.\right] \label{K2}\\
& {\cal M}_3^{(2)}(r') = \frac{r'^2}{16} \log(k r_2) \left(r'^2-4\tau^2+ 4(2\tau^2-r'^2) \log(k r') \right) + \frac{1}{16}r'^2(2\tau^2-r'^2) \left[ Li_2\left( e^{i\alpha}\frac{r_1}{r_2 } \right)+c.c.\right] \label{K3}\\
&{\cal M}_4^{(2)}(r') = \frac{r'^2}{32} \left( 8\tau^2-r'^2+4 \log(k r')\left(r'^2-4\tau^2+2(2\tau^2-r'^2)\log(k r')\right) \right)\notag\\&
\hspace{0.5in}+\frac{1}{32}\Bigg[ \left( r'^2-e^{i\alpha}r_1 r_2 \right) \left( r'^2+e^{i\alpha}r_1 r_2-4\tau^2 \right)\log\left( 1-\frac{e^{i\alpha}r_1 r_2 }{r'^2} \right) \notag\\&
\hspace{0.5in} -e^{i\alpha}r_1 r_2 \left(r'^2+\left(e^{i\alpha}r_1 r_2 -4\tau^2 \right) \log(r'^2) \right) 
+2r'^2 \left( 2 \tau^2-r'^2 \right ) Li_2\left( \frac{e^{i\alpha}r_1 r_2}{r'^2} \right) +c.c. \Bigg ] \label{K4}
\end{align}
\end{subequations}

In addition to the explicit form of the integrals ${\cal M}^{(0)}$ (equation (\ref{I})) and ${\cal M}^{(2)}$ (equation (\ref{K})) that we have calculated in this Appendix, we find it useful to introduce tables \ref{Il}, \ref{Kl} and \ref{IK}. These tables display the behavior of ${\cal M}^{(0)}$ and ${\cal M}^{(2)}$ in the six possible kinematical regions associated with the collision (see figure \ref{re}) and will be useful in the analysis of the results of section \ref{VIC}.

\begin{table}[h]
\caption{}
\centering
\begin{tabular}{c|ccc}
\hline\hline
region &ordering & Leading log behavior of ${\cal M}^{(0)}$& Sub-leading log behavior of ${\cal M}^{(0)}$  \\
\hline\hline
I & $\tau<r_1<r_2 $ &  $\frac{1}{2}\tau^2 \log(kr_1)\log(k r_2) $ &   0 \\
II &  $r_1<\tau<r_2 $ & $ \frac{1}{2}\tau^2 \log(kr_2)\log(k \tau)$ & $\frac{1}{4}(r_1^2-\tau^2) \log(k r_2) $ \\
III &  $r_1<r_2< \tau $  &  $\frac{1}{2}\tau^2 \log^2(k \tau)$ &    $\frac{1}{4} r_1^2\log(kr_2) +\frac{1}{4} r_2^2\log(kr_2) -\frac{1}{2}\tau^2 \log(k \tau) $  \\
III$^{'}$ & $r_2<r_1< \tau $  & $ \frac{1}{2}\tau^2 \log^2(k \tau)$  &    $\frac{1}{4} r_1^2\log(kr_1) +\frac{1}{4} r_2^2\log(kr_1) -\frac{1}{2}\tau^2 \log(k \tau) $ \\
II$^{'}$ &   $r_2<\tau<r_1 $ &  $\frac{1}{2}\tau^2 \log(kr_1)\log(k \tau) $ & $\frac{1}{4}(r_2^2-\tau^2) \log(k r_1) $ \\
I$^{'}$ &  $\tau<r_2<r_1 $  &  $\frac{1}{2}\tau^2 \log(kr_1)\log(k r_2) $ & 0 \\
\hline\hline
\end{tabular}
\label{Il}
\end{table}

\begin{table}[h]
\caption{}
\centering
\begin{tabular}{c|ccc}
\hline\hline
region &ordering & Leading log behavior of ${\cal M}^{(2)}$& Sub-leading log behavior of ${\cal M}^{(2)}$ \\
\hline\hline
I & $\tau<r_1<r_2 $ &  $\frac{1}{4}\tau^4 \log(kr_1)\log(k r_2) $ & 0\\
II &  $r_1<\tau<r_2 $ &   $ \frac{1}{4}\tau^4 \log(kr_2)\log(k \tau)$ & $ -\frac{1}{16}(r_1^4 - 4r_1^2\tau^2+3 \tau^4)\log(k r_2)$  \\
III &  $r_1<r_2< \tau $  &   $\frac{1}{4}\tau^4 \log^2(k \tau)$ &$-\frac{1}{16}
\Big[(r_1^4+r_2^4-4(r_1^2+r_2^2)\tau^2)\log(kr_2)+6\tau^4\log(k\tau)$ \\
&&& $-2r_1r_2\left(-4\tau^2\cos(\alpha)+r_1r_2 \cos(2\alpha)  \right)\log(\frac{r_2}{\tau}) \Big]$ \\
III$^{'}$ & $r_2<r_1< \tau $   &   $ \frac{1}{4}\tau^4 \log^2(k \tau)$ &   $-\frac{1}{16}
\Big[(r_1^4+r_2^4-4(r_1^2+r_2^2)\tau^2)\log(kr_1)+6\tau^4\log(k\tau)$ \\
&&& $-2r_1r_2\left(-4\tau^2\cos(\alpha)+r_1r_2 \cos(2\alpha)  \right)\log(\frac{r_1}{\tau}) \Big]$ \\
II$^{'}$ &   $r_2<\tau<r_1 $ &  $\frac{1}{4}\tau^4 \log(kr_1)\log(k \tau) $ &  $ -\frac{1}{16}(r_2^4 - 4r_2^2\tau^2+3 \tau^4)\log(k r_1)$ \\
I$^{'}$ &  $\tau<r_2<r_1 $   &  $\frac{1}{4}\tau^4 \log(kr_1)\log(k r_2) $ & 0 \\
\hline\hline
\end{tabular}
\label{Kl}
\end{table}

\begin{table}[h]
\caption{}
\centering
\begin{tabular}{c|cccc}
\hline\hline
region &ordering &Non-log behavior of ${\cal M}^{(0)}$&Non-log behavior of ${\cal M}^{(2)}$  \\
\hline\hline
I & $\tau<r_1<r_2 $ &$\frac{1}{8}\cos(\alpha)\frac{\tau^4}{r_1 r_2}+O(\frac{\tau^6}{r_1^2r_2^2})$  & $\frac{1}{24}\cos(\alpha)\frac{\tau^6}{r_1 r_2}+O(\frac{\tau^8}{r_1^2r_2^2})$ \\
III & $r_1<r_2< \tau $  &  $\frac{1}{4}\tau^2+O(r_1r_2\log(\frac{r_1r_2}{\tau^2}))$ & $\frac{7}{32}\tau^4+\tau^2\left( O(r_1^2)+O(r_1r_2)+O(r_2^2)\right)$ \\
III$^{'}$ &  $r_2<r_1< \tau $    &   $\frac{1}{4}\tau^2+O(r_1r_2\log(\frac{r_1r_2}{\tau^2}))$ & $\frac{7}{32}\tau^4+\tau^2\left( O(r_1^2)+O(r_1r_2)+O(r_2^2)\right)$\\
I$^{'}$ &     $\tau<r_2<r_1 $    & $\frac{1}{8}\cos(\alpha)\frac{\tau^4}{r_1 r_2}+O(\frac{\tau^6}{r_1^2r_2^2})$   & $\frac{1}{24}\cos(\alpha)\frac{\tau^6}{r_1 r_2}+O(\frac{\tau^8}{r_1^2r_2^2})$  \\
\hline\hline
\end{tabular}
\label{IK}
\end{table}

\providecommand{\href}[2]{#2}\begingroup\raggedright\endgroup
\end{document}